\documentclass[a4paper,11pt]{article}
\pdfoutput=1 \usepackage{jheppub_2} 
\usepackage{booktabs}
\usepackage[T1]{fontenc} 
\usepackage{graphicx, multirow,soul,url,amsmath,amsfonts,amssymb,mathrsfs,amsfonts}
\usepackage[all]{hypcap}
\usepackage{url}
\usepackage{bbm}
\usepackage{changepage}
\usepackage[section]{placeins}
\usepackage{cancel,ulem}
\usepackage{sidecap}
\usepackage{graphicx}
\usepackage{slashed}
\usepackage[dvipsnames]{xcolor}
\usepackage{multicol, blindtext}
\usepackage[section]{placeins}
\usepackage[version=3]{mhchem}
\usepackage{caption}
\expandafter\def\csname ver@subfig.sty\endcsname{}
\usepackage{svg}
\usepackage{lipsum}
\usepackage{hyperref}
\usepackage{footnote}
\usepackage{bbold}
\usepackage{amsbsy}
\usepackage[export]{adjustbox}
\usepackage{units}

\newcommand{\equaref}[1]{Eq.~(\ref{#1})}

\newcommand{\figref}[1]{Fig.~\ref{#1}}
\newcommand{\figsref}[2]{Figs.~\ref{#1}~and~\ref{#2}}
\newcommand{\secref}[1]{Section~\ref{#1}}

\newcommand{\tabref}[1]{Table~\ref{#1}}

\usepackage{float}
\usepackage{pifont}
\restylefloat{table}
\usepackage{lipsum}

\def\thefootnote{\fnsymbol{footnote}}
\allowdisplaybreaks

\newcommand{\be}{\begin{equation}}
\newcommand{\ee}{\end{equation}}
\newcommand{\bea}{\begin{eqnarray}}
\newcommand{\eea}{\end{eqnarray}}
\newcommand{\ba}{\begin{aligned}}
\newcommand{\ea}{\end{aligned}}

\newcommand{\bq}{\begin{eqnarray}}
\newcommand{\nq}{\end{eqnarray}}

\usepackage{subfig}

\title{\bf A Predictive and Testable Unified Theory of Fermion Masses, Mixing and Leptogenesis}
\author[a]{Bowen Fu,}
\author[a]{Stephen F. King,}
\affiliation[a]{School of Physics and Astronomy, University of Southampton, Southampton, SO17 1BJ, U.K.}
\author[b]{Luca Marsili,}
\affiliation[b]{Dipartimento di Fisica e Astronomia, Universit\`a di Bologna, via Irnerio 46, 40126 Bologna, Italy}
\author[b,c,d]{Silvia Pascoli,}
\affiliation[c]{INFN, Sezione di Bologna, viale Berti Pichat 6/2, 40127 Bologna, Italy}
\affiliation[d]{CERN, Theoretical Physics Department, Geneva, Switzerland}
\author[e]{Jessica Turner}
\affiliation[e]{Institute for Particle Physics Phenomenology, Department of Physics, Durham University, Durham DH1 3LE, U.K.}
\author[f,g]{and Ye-Ling Zhou}
\affiliation[f]{School of Fundamental Physics and Mathematical Sciences, Hangzhou Institute for Advanced
Study, UCAS, Hangzhou, China}
\affiliation[g]{International Centre for Theoretical Physics Asia-Pacific, Beijing/Hangzhou, China}

\emailAdd{b.fu@soton.ac.uk}
\emailAdd{s.f.king@soton.ac.uk}
\emailAdd{luca.marsili@studio.unibo.it}
\emailAdd{silvia.pascoli@unibo.it}
\emailAdd{jessica.turner@durham.ac.uk}
\emailAdd{zhouyeling@ucas.ac.cn}

\abstract{
We consider a minimal non-supersymmetric $SO(10)$ Grand Unified Theory (GUT) model that can reproduce the observed 
fermionic masses and mixing parameters of the Standard Model. We calculate the scales of spontaneous symmetry breaking from the GUT to the Standard Model gauge group using two-loop renormalisation group equations. This procedure determines the proton decay rate and the scale of $U(1)_{B-L}$ breaking, which generates cosmic strings and the right-handed neutrino mass scales. Consequently, the regions of parameter space where thermal leptogenesis is viable are identified and correlated with the fermion masses and mixing, the neutrinoless double beta decay rate, the proton decay rate, and the gravitational wave signal resulting from the network of cosmic strings. We demonstrate that this framework, which can explain the Standard Model fermion masses and mixing and the observed baryon asymmetry, will be highly constrained by the
next generation of gravitational wave detectors and neutrino oscillation experiments which will also constrain the proton lifetime.}

\preprint{
\begin{flushleft}{
 IPPP/22/57\,,  CERN-TH-2022-141}
\end{flushleft}
}


\begin{document}

\thispagestyle{empty}
\def\thefootnote{\fnsymbol{footnote}}
\setcounter{footnote}{1}

\setcounter{page}{0}
\maketitle
\vspace{-1cm}
\flushbottom

\def\thefootnote{\arabic{footnote}}
\setcounter{footnote}{0}

\section{Introduction}
Grand Unified Theories (GUTs) have long been an attractive framework for unifying the non-gravitational interactions. The minimal option, which can predict neutrino masses and mixing, uses the gauge group $SO(10)$. Several well-studied symmetries can be embedded in $SO(10)$, 
including $SU(5)$ \cite{Georgi:1972cj}, flipped $SU(5) \times U(1)$~\cite{Barr:1981qv,Derendinger:1983aj,DeRujula:1980qc,Antoniadis:1989zy} and the Pati-Salam model $SU(4)_c \times SU(2)_L \times SU(2)_R$~\cite{Pati:1973uk}. Thanks to this rich structure, there are many possible symmetry-breaking chains from $SO(10)$ down to the Standard Model (SM) gauge group, $G_{SM}$, most of them via the Pati-Salam symmetry \cite{Jeannerot:2003qv}. An appealing feature of an intermediate Pati-Salam symmetry in non-supersymmetric GUTs is that gauge unification can be achieved, and there is an intermediate $U(1)_{B-L}$ subgroup which is spontaneously broken, generating right-handed neutrino masses. In addition to inducing light neutrino masses via the seesaw mechanism, the CP-violating and out-of-equilibrium decays of the right-handed neutrinos can produce the observed matter-antimatter asymmetry via thermal leptogenesis \cite{Fukugita:1986hr}. Moreover, the $U(1)_{B-L}$ symmetry breaking can also generate cosmic strings in the early Universe, which can intercommute and emit gravitational radiation forming a stochastic gravitational wave (GW) background that future GW interferometers can test. 

The connection between GUTs and gravitational waves has been studied in \cite{Buchmuller:2019gfy} where the simple breaking pattern $SO(10)\to G_{SM}\times U(1)_{B-L}\to G_{SM}$ was shown to be consistent with inflation, leptogenesis, and dark matter, while the $U(1)_{B-L}$ symmetry breaking generates cosmic strings. The connection between high-scale thermal leptogenesis and GWs was also pointed in \cite{Dror:2019syi} where it was assumed that the $U(1)_{B-L}$ breaking scale is the same as the seesaw and leptogenesis scales.
In Ref.~\cite{King:2020hyd}, we highlighted the complementarity between proton decay and gravitational wave signals from cosmic strings as a powerful method of probing GUTs. Subsequently, in Ref.~\cite{King:2021gmj}, we studied all possible non-supersymmetric $SO(10)$ symmetry-breaking chains. We performed a comprehensive renormalisation group (RG) analysis to find the correlations between the proton decay rate and the GW signal. We also identified which chains survived the current non-observation of both proton decay and GWs and could be tested by future neutrino and GW experiments.

In this paper, we go beyond these works by providing a detailed study on a specific $SO(10)$ breaking chain that provides unification and predicts a proton decay width via the channel $p \to \pi^0 e^+$, consistent with the experimental bound of the Super-Kamiokande (Super-K) \cite{Super-Kamiokande:2020wjk} and can be fully tested by future proton decay searches of Hyper-K~\cite{King:2021gmj}. Further, this breaking chain generates cosmic strings at the lowest intermediate scale, $M_1\sim 10^{13}$ GeV. A GW background generated by such a string network is just around the corner and may be even already hinted at by recent observations in PTA experiments, including NANOGrav \cite{Arzoumanian:2020vkk}, PPTA \cite{Goncharov:2021oub}, EPTA \cite{Chen:2021rqp} and IPTA \cite{Antoniadis:2022pcn}. We determine the minimal necessary particle content to induce the pattern of breaking and perform an RG analysis and numerical fit of our model to SM data to postdict the fermion masses and mixing, including the mass scales of the right-handed neutrinos. As this procedure determines the scales of symmetry breaking of our model and the masses of the right-handed neutrinos, the matter-antimatter asymmetry associated with thermal leptogenesis is predicted. We then show that successful leptogenesis can occur in the regions of the model parameter space consistent with SM fermion masses and mixing and can be correlated with a GW signal and proton decay. Compared with \cite{Dror:2019syi}, such an approach allows us to go beyond generic considerations and instead to quantitatively account for the hierarchy between the leptogenesis and see-saw scales, as well as with the $U(1)_{B-L}$ breaking scale, thanks to the constraints imposed by reproducing the low energy data. The latter scale is of particular interest since pulsar timing arrays such as PPTA \cite{Manchester:2012za} and NANOGrav \cite{Arzoumanian:2018saf} are sensitive to the predicted GW signals while future large-scale neutrino experiment, Hyper-Kamiokande (Hyper-K) \cite{Hyper-Kamiokande:2018ofw}, will be able to probe the predicted proton decay rate of this model. The correlation between these two observables will be a crucial test of our GUT model, and such methodology can be applied to other GUT models, presenting a new avenue to try to unveil the physics at very high scales.

This paper is organised as follows: in \secref{sec:framework}, we discuss the GUT symmetry breaking pattern and the particle content of our model, including fermionic and Higgs representations of the GUT and our RG running procedure. In \secref{sec:fandf}, we discuss how we relate our model to the quark lepton data, our fitting procedure and the ensuing results. In \secref{sec:leptogenesis}, we discuss the basics of non-resonant thermal leptogenesis and how we determine the baryon asymmetry produced from the successful points in the model parameter space and in \secref{sec:testability}, we demonstrate that the regions of the model parameter space that yield successful leptogenesis and fermionic masses and mixing will be associated with a GW signal. Finally, we summarise and discuss in \secref{sec:conclusion}.
As a case study, we consider a benchmark point (referred to as BP1 throughout) and discuss how it satisfies all these experimental constraints in each section.
\section{The framework}\label{sec:framework}
We focus on a breaking chain (classified as chain III4 of type (c) in Ref.~\cite{King:2021gmj}) that is of particular interest as it is currently allowed and predicts a  proton decay rate testable by Hyper-K. We discuss the breaking chain's
 matter content and gauge unification in this section.
\subsection{Symmetry breaking of $SO(10)$}\label{sec:breaking}
We study the following breaking chain with three intermediate symmetries 
($G_{3}$, $G_{2}$, and $G_{1}$):
\begin{eqnarray}
& SO(10) \nonumber\\
&\hspace{20mm}{\bf 54}~\Big\downarrow~\text{broken at} ~ M_X \nonumber\\
&G_3 \equiv SU(4) \times SU(2)_L \times SU(2)_R \times Z_2^C \nonumber\\
&\hspace{16.5mm}{\bf 210}~\Big\downarrow~\text{broken at} ~ M_3 \nonumber\\
&G_2 \equiv SU(3)_c \times SU(2)_L \times SU(2)_R \times U(1)_{X} \times Z_2^C \nonumber\\
&\hspace{18.5mm}{\bf 45}~\Big\downarrow~\text{broken at} ~ M_2 \nonumber\\
&G_1 \equiv SU(3)_c \times SU(2)_L \times SU(2)_R \times U(1)_{X} \nonumber\\
&\hspace{16mm}{\overline{\bf 126}}~\Big\downarrow~\text{broken at} ~ M_1 \nonumber\\ 
&G_{\rm SM} \equiv SU(3)_c \times SU(2)_L \times U(1)_Y \,.
\end{eqnarray}
The boldface number beside the arrow indicates the Higgs representation of SO(10), triggering the symmetry breaking. 
In this work, we follow the same convention as Ref.~\cite{King:2021gmj} where the GUT symmetry breaking scale is denoted as $M_X$ and the mass scale of the subsequent breaking of the group $G_I$ (for $I=1,2,3$) is denoted as $M_I$.\footnote{However, we change the notation of the running energy scale to from $\mu$ to $Q$ as the string tension is often denoted as $\mu$.} All particles, except the gauge fields of the model, are listed in Table~\ref{tab:particle_content}. We note that $Z_2^C$ refers to the parity symmetry between left and right conjugation ($L \leftrightarrow R^c$, where $c$ indicates charge conjugation) and that 
$U(1)_X$ is identical to the $U(1)_{B-L}$ symmetry with the charge correlated via $X = \sqrt{\frac{3}{2}}(\frac{B-L}{2})$. The correlations between $U(1)$ charges are given by $Y = \sqrt{\frac{3}{5}} \left( I_{3R} + \frac{B-L}{2} \right)$, where $I_{3R}$ is the isospin in $SU(2)_R$.

\begin{table}[t!] 
\begin{center}
\begin{tabular}{|c | c | l |}
\hline \hline 
 & Multiplet & Role in the model
\\\hline
Fermions & ${\bf 16}$ & Contains all SM fermions and RH neutrinos \\\hline
 &${\bf 10}$ & Generates fermion masses \\
 &${\bf 45}$ & Triggers intermediate symmetry breaking \\
Higgses &${\bf 54}$ & Triggers GUT symmetry breaking \\
 &${\bf 120}$ & Generates fermion masses \\
 &$\overline{\bf 126}$ & Generates fermion masses \& intermediate symmetry breaking \\
 &${\bf 210}$ & Triggers intermediate symmetry breaking \\
\hline \hline
\end{tabular}
\end{center}
\caption{{The $SO(10)$ representations of the fermion and Higgs particles of our $SO(10)$ GUT model and their roles.} \label{tab:particle_content}}
\end{table}

To achieve each step of symmetry breaking, i.e., $SO(10) \to G_3 \to G_2 \to G_1$, we include three Higgs multiplets, ${\bf 54}$, ${\bf 210}$, and ${\bf 45}$ of $SO(10)$, respectively. These Higgs fields spontaneously break the GUT symmetry as follows:
\begin{itemize}
\item 
${\bf 54}$ contains a parity-even singlet $({\bf 1}, {\bf 1},{\bf 1})$ of $G_3 \equiv SU(4)_c \times SU(2)_L \times SU(2)_R$ where each entry in the bracket $({\bf r_1}, {\bf r_2}, \cdots)$ refers to the field representation transforming in the group $G \equiv H_1 \times H_2 \times \cdots$. Once $({\bf 1}, {\bf 1},{\bf 1})$ gains a non-trivial vacuum expectation value (VEV) at scale $M_X$, $SO(10)$ is spontaneously broken to $G_3$. 
\item
In $G_3$, ${\bf 210}$ can be decomposed to $({\bf 15}, {\bf 1},{\bf 1})_1$ of $G_3$, which is further decomposed into a parity-even and trivial singlet $({\bf 1}, {\bf 1},{\bf 1}, 0)_1$ of $SU(3)_c \times SU(2)_L \times SU(2)_R \times U(1)_X$, where the last entry in the bracket is the field charge in the $U(1)$ symmetry and the subscript is used to distinguish from another field with the same representation discussed below. The VEV of this singlet breaks $G_3$ to $G_2$ at scale $M_3$. 
\item
The breaking of $G_2$ to $G_1$ is realised via a $\bf 45$ of $SO(10)$, which decomposed into
$({\bf 15}, {\bf 1},{\bf 1})_2$ of $G_3$ and a further $({\bf 1}, {\bf 1},{\bf 1}, 0)_2$ of $G_2$ and $G_1$.
This singlet is parity-odd, and its VEV induces the breaking of $G_2 \to G_1$ at scale $M_{2}$.
\item Finally, the breaking of $G_1 \to G_{\rm SM}$ at scale $M_1$ is provided by $\overline{\bf 126}$, which is decomposed into a triplet $({\bf 1}, {\bf 1}, {\bf 3}, -1)$ of $SU(3)_c \times SU(2)_L \times SU(2)_R \times U(1)_X$ and to a further singlet, $({\bf 1}, {\bf 1}, 0)$, of $G_{\rm SM}$. This singlet field, denoted as $\phi_{S}$, provides mass to the three right-handed neutrinos.
\end{itemize}
We summarise the decomposition of Higgses, which triggers the breaking of $SO(10)$ and intermediate symmetries, in Table~\ref{tab:decomposition_H_breaking}. 

\begin{table}[tbp] 
\begin{center}
\begin{tabular}{| c | c | c | c | c |}
\hline \hline 
$SO(10)$ \quad
& ${\bf 54}$ &
 ${\bf 210}$ &
 ${\bf 45}$ &
 ${\overline{\bf 126}}$
\\\hline
$G_3$ & $({\bf 1}, {\bf 1},{\bf 1})$ &
$({\bf 15}, {\bf 1},{\bf 1})_1$ 
& $({\bf 15}, {\bf 1},{\bf 1})_2$
& $({\bf 10}, {\bf 1},{\bf 3}) + (\overline{\bf 10}, {\bf 3},{\bf 1})$ \\\hline
$G_2$ & --
&
$({\bf 1}, {\bf 1},{\bf 1},0)_1$ &
$({\bf 1}, {\bf 1},{\bf 1}, 0)_2$
& $({\bf 1}, {\bf 1}, {\bf 3}, -1) + ({\bf 1}, {\bf 3}, {\bf 1}, 1)$ \\\hline
$G_1$ & --
& --
&
$({\bf 1}, {\bf 1},{\bf 1}, 0)_2$
& $({\bf 1}, {\bf 1}, {\bf 3}, -1)$ \\\hline
$G_{\rm SM}$ & --
& --
& --
&
$({\bf 1}, {\bf 1}, 0)_S$ \\
\hline \hline
\end{tabular}
\end{center}
\caption{{Decomposition of the Higgses which induce spontaneous symmetry breaking at each step of the breaking chain. Each Higgs (from left to right) is eventually decomposed to a singlet whose non-vanishing VEV preserves the symmetry $G_I$ (for $I = 3,2,1,{\rm SM}$) in the same row but breaks larger symmetries. The subscript distinguishes different fields of the same representation.} \label{tab:decomposition_H_breaking}}
\end{table}

\subsection{Matter Field Decomposition and Fermion Masses}\label{sec:matter}
\begin{table}[t!] 
\begin{center}
\begin{tabular}{| c | c |}
\hline \hline 
$SO(10)$ & ${\bf 16}$
\\\hline & \\[-5mm]
$G_3$ &
$({\bf 4}, {\bf 2}, {\bf 1})_L + (\overline{\bf 4}, {\bf 1}, {\bf 2})_{R^c}$ \\[-5mm] & \\\hline
& \\[-5mm] & \\[-5mm]
\multirow{2}*{$G_2$} &
$({\bf 3}, {\bf 2},{\bf 1}, 1/6)_{Q_L} + (\overline{\bf 3}, {\bf 1},{\bf 2}, -1/6)_{Q_R^c}$ \\
& $+ ({\bf 1}, {\bf 2}, {\bf 1}, -1/2)_{l_L} + ({\bf 1}, {\bf 1},{\bf 2}, 1/2)_{l_R^c}$ \\[1mm]\hline
& \\[-5mm]
\multirow{2}*{$G_1$} &
$({\bf 3}, {\bf 2},{\bf 1}, 1/6)_{Q_L} + (\overline{\bf 3}, {\bf 1},{\bf 2}, -1/6)_{Q_R^c}$ \\ 
& $+ ({\bf 1}, {\bf 2}, {\bf 1}, -1/2)_{l_L} + ({\bf 1}, {\bf 1},{\bf 2}, 1/2)_{l_R^c}$ \\[1mm]\hline
& \\[-5mm]
\multirow{2}*{$G_{\rm SM}$} &
$({\bf 3}, {\bf 2},1/6)_{Q_L} + (\overline{\bf 3}, {\bf 1},-2/3)_{u_R^c} + (\overline{\bf 3}, {\bf 1},1/3)_{d_R^c}$ \\
& $+({\bf 1}, {\bf 2}, -1/2)_{l_L}\!+\!({\bf 1}, {\bf 1}, 0)_{\nu_R^c} + ({\bf 1}, {\bf 1},1)_{e_R^c}$
\\[1mm]
\hline \hline
\end{tabular}
\end{center}
\caption{{Decomposition of the matter multiplet ${\bf 16}$ in each step of the breaking chain.} \label{tab:decomposition_f}}
\end{table}
In order to assess if our model can predict the measured fermionic masses and mixing, it is important to understand the matter content of the 
breaking chain. Fermions are arranged as a ${\bf 16}$ of $SO(10)$ and follow the decomposition given in Table~\ref{tab:decomposition_f} where $L$ ($R$) denote the left-handed (right-handed) fermions of $G_3$ which contains the SM left-handed (right-handed) fermions where $Q_{L (R)}$ and $\ell_{L (R)}$ are the quark and leptonic $\mathrm{SU}(2)_{L(R)}$ doublets, respectively, and $u_{R}, d_{R}, e_{R}$, and $\nu_{R}$ are the quark and lepton $\mathrm{SU}(2)_{L}$ singlets, respectively. 

Three Higgs multiplets, ${\bf 10}$, $\overline{\bf 126}$ and ${\bf 120}$, are required to generate the Standard Model fermion masses. Compared to Ref.~\cite{King:2021gmj}, where we considered a minimal survival hypothesis \cite{delAguila:1980qag}, we include one additional Higgs (${\bf 120}$) which 
is required to generate all fermion mass spectra, mixing angles, and CP-violating phases in the quark and lepton sectors.
 Here, $\overline{\bf 126}$ is the same Higgs used in the breaking $G_1 \to G_{\rm SM}$. For this breaking chain, we list decompositions of Higgs, which are responsible for mass generation in \tabref{tab:decomposition_H_breaking}. 
 
Applying this decomposition, we have two $({\bf 1}, {\bf 2}, {\bf 2})$ and two $({\bf 15}, {\bf 2}, {\bf 2})$ multiplets of $G_3$ after the $SO(10)$ breaking. These multiplets are composed of four bi-doublets, $({\bf 1}, {\bf 2}, {\bf 2}, 0)$, of $G_2$ and $G_1$. After $SU(2)_R$ is broken below $M_1$, each bi-doublet contains two electroweak doublets $({\bf 1}, {\bf 2}, \mp 1/2)$, 
and eventually, we arrive at the eight electroweak doublets of the model which we denote as ${h_i} = \{\tilde{h}_{\bf 10}^u, \tilde{h}_{\overline{\bf 126}}^u, \tilde{h}_{\bf 120}^u, \tilde{h}_{\bf 120}^{u'}, h_{\bf 10}^d, h_{\overline{\bf 126}}^d, h_{\bf 120}^d, h_{\bf 120}^{d'} \}$, where $\tilde{h}_{\bf 10}^u = i \sigma_2 (h_{\bf 10}^{u})^*$. These field decompositions introduce particles beyond the SM spectrum and 
 may contribute to the renormalisation group running behaviour of the gauge coefficients. However, we reduce their redundancy in the following way: For scale $Q$ which varies in the range $M_X > Q > M_3$, where $G_3$ is preserved, the two decomposed $({\bf 1}, {\bf 2}, {\bf 2})$'s can mix and we assume that the heavy one gains a mass $\sim M_X$ and thus decouples at scales below $M_X$. The same assumption applies to the other two $({\bf 1}, {\bf 2}, {\bf 2})$'s. Using these assumptions, we have two bi-doublets $({\bf 1}, {\bf 2}, {\bf 2}, 0)$ at the scale $M_3 > Q > M_2$ and $M_2 > Q > M_1$, where $G_2$ and $G_1$ are preserved, respectively. We retain them as the physically relevant degrees of freedom in this range of scales, following the logic of Ref.~\cite{King:2021gmj}. Four electroweak doublets remain at energies below $M_1$ but above the electroweak scale. Naively, one can assume all massive states are sufficiently heavy that they decouple at scale $M_1$, except for the lightest electroweak doublet, which is the SM Higgs and should be massless before electroweak symmetry breaking. Without loss of generality, we can write these Higgses as superpositions of mass eigenstates, $\hat{h}_i = \sum_j V_{ij} h_j$, with $h_{\rm SM} \equiv \hat{h}_1$, where $V$ is a unitary matrix and the heavy doublets that decouple at $M_X$ have also been taken into account. With this treatment, all physical degrees of freedom present at the relevant scale are the same as those of chain III4 in Ref.~\cite{King:2021gmj}. 
\begin{table}[tbp] 
\begin{center}
\begin{tabular}{| c | c | c | c |}
\hline \hline 
$SO(10)$ & ${\bf 10}$ & ${\overline{\bf 126}}$ & ${\bf 120}$ 
\\\hline
\multirow{2}*{$G_3$} & $({\bf 1}, {\bf 2},{\bf 2})_1$ &
$({\bf 15}, {\bf 2},{\bf 2})_1$ &
$({\bf 1}, {\bf 2},{\bf 2})_2 + ({\bf 15}, {\bf 2},{\bf 2})_2$ \\
&& $+({\bf 10}, {\bf 1},{\bf 3}) + (\overline{\bf 10}, {\bf 3},{\bf 1})$ &
 \\\hline
\multirow{2}*{$G_2$} &
$({\bf 1}, {\bf 2},{\bf 2}, 0)_1$ &
$({\bf 1}, {\bf 2},{\bf 2},0)_2$ &
$({\bf 1}, {\bf 2},{\bf 2}, 0)_{3,4}$ \\
&& $+ ({\bf 1}, {\bf 1}, {\bf 3}, -1) + ({\bf 1}, {\bf 3}, {\bf 1}, 1)$ & 
\\\hline
\multirow{2}*{$G_1$} & 
$({\bf 1}, {\bf 2},{\bf 2}, 0)_1$ &
$({\bf 1}, {\bf 2},{\bf 2},0)_2$ &
$({\bf 1}, {\bf 2},{\bf 2}, 0)_{3,4}$ \\
&& $+ ({\bf 1}, {\bf 1}, {\bf 3}, -1)$ & \\\hline
\multirow{3}*{$G_{\rm SM}$} & $({\bf 1}, {\bf 2}, - 1/2)_{h_{\bf 10}^{u}}$ &
$({\bf 1}, {\bf 2}, - 1/2)_{h_{\overline{\bf 126}}^{u}}$ &
$({\bf 1}, {\bf 2}, - 1/2)_{h_{\bf 120}^{u}, {h_{\bf 120}^{u'}}}$ \\
&
$+ ({\bf 1}, {\bf 2}, + 1/2)_{h_{\bf 10}^{d}}$ &
$+ ({\bf 1}, {\bf 2}, + 1/2)_{h_{\overline{\bf 126}}^{d}}$ &
$+ ({\bf 1}, {\bf 2}, + 1/2)_{h_{\bf 120}^{d}, {h_{\bf 120}^{d'}}}$ \\
&& $+ ({\bf 1}, {\bf 1},0)_S$ & \\
\hline \hline
\end{tabular}
\end{center}
\caption{Decomposition of Higgses responsible for the fermion mass generation. $\overline{\bf 126}$ is the same Higgs as shown in Table~\ref{tab:decomposition_H_breaking} and it is responsible for both the breaking $G_1 \to G_{\rm SM}$ and right-handed neutrino mass generation. $({\bf 1}, {\bf 1},0)_S$ is the same singlet given in Table~\ref{tab:decomposition_H_breaking}. }\label{tab:decomposition_H_mass}
\end{table}
For the second Higgs multiplet, $\overline{\bf 126}$, we retain another decomposed representation $({\bf 10}, {\bf 3}, {\bf 1})$ of $G_1$, which contains a $SU(2)_L$ triplet, $({\bf 1}, {\bf 1}, {\bf 3}, -1)$, of $G_2$ and $G_1$ which contains the singlet $S \sim ({\bf 1}, {\bf 1},0)$ of $G_{\rm SM}$ that is important not only in its role in symmetry breaking, but also in the generation of neutrino masses. 
$(\overline{\bf 10}, {\bf 3}, {\bf 1})$ is retained due to the requirement of left-right parity symmetry, $Z_2^C$, and it is decomposed to a $({\bf 1}, {\bf 3}, {\bf 1}, 1)$ of $G_2$. After $G_2$ breaking, i.e., the breaking of the left-right parity symmetry, we assume that this particle decouples.

In the Yukawa sector, couplings above the GUT scale are given by
\begin{eqnarray}\label{eq:yuk}
Y_{\bf 10}^* \, {\bf 16} \cdot {\bf 16} \cdot {\bf 10} + Y_{\overline{\bf 126}}^* \, {\bf 16} \cdot {\bf 16} \cdot \overline{\bf 126} + Y_{\bf 120}^* \, {\bf 16} \cdot {\bf 16} \cdot {\bf 120} + {\rm h.c.}\,,
\end{eqnarray}
where the asterisk denotes complex conjugation.
Considering the flavour indices, $Y_{\bf 10}$ and $Y_{\overline{\bf 126}}$ are in general complex $3\times 3$ symmetric matrices and $Y_{\bf 120}$ is an antisymmetric matrix. 
In the non-SUSY case, two further couplings ${\bf 16} \cdot {\bf 16} \cdot {\bf 10}^*$ and ${\bf 16} \cdot {\bf 16} \cdot {\bf 120}^*$ are allowed by the gauge symmetry; however, we forbid them by imposing an additional Peccei–Quinn $U(1)$ symmetry \cite{Peccei:1977hh} as described in \cite{Joshipura:2011nn,Bajc:2005zf,Babu:1992ia}.
After the final symmetry is broken to $G_{\rm SM}$, the above Yukawa terms generate the following SM fermion mass terms in the left-right convention:
\be
\ba
Y_{\bf 10} & \Big[ ( \overline{Q} u_R + \overline{L} \nu_R ) h_{\bf 10}^u + (\overline{Q} d_R + \overline{L} e_R) h_{\bf 10}^d \Big]
+ 
\displaystyle\frac{1}{\sqrt{3}}Y_{\overline{\bf 126}} \left[( \overline{Q} u_R -3 \overline{L} \nu_R) h_{\overline{\bf 126}}^u + (\overline{Q} d_R -3 \overline{L} e_R) h_{\overline{\bf 126}}^d \right] \\
+ Y_{\bf 120} &\Big[ ( \overline{Q} u_R + \overline{L} \nu_R) h_{\bf 120}^u + (\overline{Q} d_R + \overline{L} e_R ) h_{\bf 120}^d + 
\frac{1}{\sqrt{3}} ( \overline{Q} u_R -3 \overline{L} \nu_R) h_{\bf 120}^{u'} + (\overline{Q} d_R -3 \overline{L} e_R ) h_{\bf 120}^{d'} \Big] + {\rm h.c.}\,
\ea
\ee
Rotating the Higgs fields to their mass basis, we derive Yukawa couplings to the SM Higgs as  
\begin{eqnarray}
Y_u \overline{Q} \, \tilde{h}_{\rm SM} \, u_R + Y_d \overline{Q} \, {h}_{\rm SM} \, d_R + 
Y_\nu \overline{L} \, \tilde{h}_{\rm SM} \, \nu_R + Y_e \overline{L} \, {h}_{\rm SM} \, e_R + {\rm h.c.}\,,
\end{eqnarray}
where  
\begin{eqnarray}
Y_u &=& Y_{\bf 10} V_{11}^* + \frac{1}{\sqrt{3}} Y_{\overline{\bf 126}} V_{12}^* + 
Y_{\bf 120} \left(V_{13}^* + \frac{1}{\sqrt{3}} V_{14}^*\right) \,, \nonumber\\
Y_d &=& Y_{\bf 10} V_{15} + \frac{1}{\sqrt{3}}Y_{\overline{\bf 126}} V_{16} +
Y_{\bf 120} \left(V_{17}+ \frac{1}{\sqrt{3}} V_{18}\right) \,, \nonumber\\
Y_\nu &=& Y_{\bf 10} V_{11}^* -\sqrt{3} Y_{\overline{\bf 126}} V_{12}^* + 
Y_{\bf 120} \left(V_{13}^* - \sqrt{3} V_{14}^*\right) \,, \nonumber\\
Y_e &=& Y_{\bf 10} V_{15} - \sqrt{3}Y_{\overline{\bf 126}} V_{16} +
Y_{\bf 120} \left(V_{17} - \sqrt{3} V_{18}\right) \,.
\end{eqnarray}
A Majorana mass term for the right-handed neutrinos is generated from the second term of \equaref{eq:yuk}:
\begin{eqnarray}
Y_{\overline{\bf 126}} 
\overline{\nu}_R \, \phi_{S} \, \nu_R^c +{\rm h.c.}\,, \label{eq:Yukawa_NS}
\end{eqnarray}
once $\phi_{S}$ acquires a VEV, $v_S$, which controls the scale of the masses:
\begin{eqnarray}
M_{\nu_R} &=& Y_{\overline{\bf 126}}\, v_{S}\,.
\end{eqnarray}
After the right-handed neutrinos decouple and electroweak symmetry is broken, the light neutrinos acquire their mass via the Type-I seesaw mechanism \cite{Mohapatra:1979ia,Gell-Mann:1979vob,Yanagida:1979as,Minkowski:1977sc}:
\begin{eqnarray}
M_\nu &=& - Y_\nu M_{\nu_R}^{-1} Y_\nu^T v_{\rm SM}^2 \,,
\end{eqnarray}
where the SM Higgs VEV is $v_{\rm SM} =175$ GeV. 
We emphasise that the electroweak singlet, $\phi_{S}$, is essential for the symmetry breaking $G_1 \to G_{\rm SM}$ and thus, its VEV determines the scale of $M_1$ and the right-handed neutrino masses. As required by perturbativity, $Y_{\overline{\bf 126}}\lesssim{\cal O}(1)$, the mass of the heaviest right-handed neutrino, $M_{N_3}$, should be not heavier than the lowest intermediate scale, $M_1$. On the other hand, neutrino oscillation experiments have given relatively precise measurements of light neutrino masses and mixing angles. These data restrict the right-handed neutrino mass spectrum via the seesaw formula, and a realistic GUT model should survive all such constraints.

\subsection{Gauge Unification} \label{sec:gauge_unifcation}
Given an arbitrary gauge symmetry $G$, which can be expressed as a product of simple Lie groups, $G= H_1 \times \cdots \times H_n$, the two-loop renormalisation group running equation for group $H_i$, for $i=1,2, \cdots$, is given by
\begin{eqnarray}
Q \frac{d\alpha_i}{d Q} = \beta_i (\alpha_i)\,,
\end{eqnarray}
where $\alpha_i =g_i^2/(4\pi)$ and the $\beta$ function is determined by the particle content of the theory:
\begin{eqnarray}
\beta_i = - \frac{1}{2\pi} \alpha_i^2 ( b_i + \frac{1}{4\pi} \sum_{j} b_{ij} \alpha_j ) \,.
\end{eqnarray}
Here, $i\in [1, \cdots, n]$ for $H_n$, $g_i$ is the gauge coefficient of $H_i$, and $b_i$ and $b_{ij}$ refer to the normalised coefficients of one- and two-loop contributions, respectively. In the following, we neglect the Yukawa contribution to the RG running equations as it gives a subdominant contribution.
Given two scales $Q_0$ and $Q$, if the conditions $Q_0< Q$ and $b_j \alpha_j(Q_0) \log(Q/Q_0) <1$ are both satisfied then an analytical solution for these equations can be obtained \cite{Bertolini:2009qj}:
\begin{eqnarray}
\alpha_i^{-1}(Q) = \alpha_i^{-1}(Q_0) - \frac{b_i}{2\pi}\log\frac{Q}{Q_0} + \sum_j \frac{b_{ij}}{4\pi b_i} \log\left(1- \frac{b_j}{2\pi} \alpha_j(Q_0) \log \frac{Q}{Q_0}\right) \,.
\end{eqnarray}
In the case that both $H_i$ and $H_j$ are non-abelian groups, the coefficients $b_i$ and $b_{ij}$ are
\begin{eqnarray}
b_i &=& - \frac{11}{3} C_2(H_i) + \frac{2}{3} \sum_F T(\psi_i) + \frac{1}{3} \sum_S T(\phi_i) \,, \nonumber\\
b_{ij} &=&
- \frac{34}{3} [C_2(H_i)]^2 \delta_{ij} + \sum_F T(\psi_i) [2 C_2(\psi_j) + \frac{10}{3} C_2(H_i) \delta_{ij}] \nonumber\\
&&+ \sum_S T(\phi_i) [4 C_2(\phi_j) + \frac{2}{3} C_2(H_i) \delta_{ij}]\,,
\end{eqnarray}
where the $\psi$ and $\phi$ indices sum over the fermions and complex scalar multiplets, respectively, and $\psi_i$ and $\phi_i$ are their representations in the group $H_i$, respectively. $C_2(R_i)$ (for $R_i=\psi_i, \phi_i$) denotes the quadratic Casimir of the representation $R_i$ in group $H_i$ and
$C_2(H_i)$ is the quadratic Casimir of the adjoint presentation of the group $H_i$. 

In particular, for $SU(N)$, $C_2(SU(N))=N$ and the quadratic Casimir of the fundamental irrep ${\bf N}$ of $SU(N)$ is given by $C_2({\bf N}) = (N^2-1)/2N$; 
for $SO(10)$, $C_2(SO(10))=8$, and the quadratic Casimir of the fundamental irrep ${\bf 10}$ of $SO(10)$ is given by $C_2({\bf 10}) = 9/2$. 
The spinor representation of $SO(10)$, ${\bf 16}$, has $C_2({\bf 16}) = 45/4$. 
$T(R_i)$ is the Dynkin index of representation $R_i$ of group $H_i$. For $SU(N)$, $T(R_i) = C_2(R_i)d(R_i)/(N^2-1)$ where $d(R_i)$ is the dimension of $R_i$.
If one of $H_j$ is a $U(1)$ symmetry, the coefficient $b_{ij}$ is obtained by replacing $C_2(R_j)$ and $T(R_j)$ with the charge square $[Q_j(R)]^2$ of the field multiplet $R$ in $U(1)_j$.
For the Abelian symmetry, $C_2(U(1)) = 0$.

Explicit values of $b_i$ and $b_{ij}$ depend on the degree of freedoms introduced by the gauge, matter and Higgs fields. The gauge fields are directly determined by the gauge symmetry in the breaking chain. In regards to the matter fields, we assume they are the minimal extension which includes all the SM fermions, i.e., minimally a ${\bf 16}$ of $SO(10)$ as in \tabref{tab:decomposition_f}. The most significant uncertainty contributing to RG running comes from the Higgs sector as one has to account for all the Higgses used to generate fermion masses and the GUT and intermediate symmetry breaking. Given the decomposition of Higgs fields in \tabref{tab:decomposition_H_breaking} and the discussion in \secref{sec:matter}, the Higgs fields included in each step of the RG running are:
{\begin{itemize}
\item For $G_1 \to G_{\rm SM}$, we include only the SM Higgs. Although we arrive at a series of electroweak doublets after field decomposition, we assume that all degrees of freedom except the SM Higgs are sufficiently heavy that they are integrated out by this breaking step and thus have a negligible effect on the RG running.  

\item For $G_2 \to G_1$, we include three Higgses in the running, two $({\bf 1}, {\bf 2}, {\bf 2}, 0)$'s and one $({\bf 1}, {\bf 1}, {\bf 3}, -1)$ of $G_1$. The former includes the SM Higgs, and the latter includes the gauge singlet $\phi_{S}$ of $G_{\rm SM}$ which is used to achieve the breaking of $G_1 \to G_{\rm SM}$ and right-handed neutrino masses.
\item 
For $G_3 \to G_2$, we include two $({\bf 1}, {\bf 2}, {\bf 2},0)$'s, $({\bf 1}, {\bf 1}, {\bf 3}, -1)$, $({\bf 1}, {\bf 3}, {\bf 1}, 1)$, and $({\bf 1}, {\bf 1}, {\bf 1},0)_2$ in the RG running. Two further Higgses are included compared to the above item as $({\bf 1}, {\bf 3}, {\bf 1}, 1)$ is required for the matter parity symmetry $Z_2^C$ and $({\bf 1}, {\bf 1}, {\bf 1}, 0)_2$ is used to break $Z_2^C$, $G_2 \to G_1$. 

\item 
For $SO(10) \to G_3$, we include $({\bf 1}, {\bf 2}, {\bf 2})$, $({\bf 15}, {\bf 2}, {\bf 2})$, $({\bf 10}, {\bf 1}, {\bf 3})$, $(\overline{\bf 10}, {\bf 3}, {\bf 1})$ and two $({\bf 15}, {\bf 1}, {\bf 1})$'s in the RG running. The former two are required to obtain the two $({\bf 1}, {\bf 2}, {\bf 2}, 0)$'s above. $({\bf 10}, {\bf 1}, {\bf 3})$ and $(\overline{\bf 10}, {\bf 3}, {\bf 1})$ are required for $({\bf 1}, {\bf 1}, {\bf 3}, -1)$ and $({\bf 1}, {\bf 3}, {\bf 1}, 1)$. One $({\bf 15}, {\bf 1}, {\bf 1})$, decomposed from ${\bf 45}$ is for $({\bf 1}, {\bf 1}, {\bf 1},0)_2$, and the other, decomposed from $210$, includes the singlet $({\bf 1}, {\bf 1}, {\bf 1},0)_1$ to achieve the breaking $G_3 \to G_2$.
\end{itemize}}
By including the above particle content in the RG running, we obtain the coefficients $b_i$ and $b_{ij}$ at the two-loop level, which we list in Table~\ref{tab:beta} and are the same as in the chain III4 of Ref.~\cite{King:2021gmj}. Although we include one more Higgs multiplet ${\bf 120}$, the contribution of induced new particles can be ignored, as explained in the previous subsection, by assuming heavy mass eigenstates heavier than the breaking scale, $M_X$ \cite{Dutta:2004hp}. 
In order to keep the treatment of the RG running economical, the scalar multiplets which are unnecessary for the breaking chain are assumed to be as massive as the SO(10) breaking scale $M_X$. Therefore, these scalars will not affect the RG running or provide threshold corrections. 

\begin{table}[tbp] 
\begin{center}
\begin{tabular}{| c l |}
\hline \hline 
$SO(10)$ & broken at $Q=M_X$ \\\hline &\\[-5mm]
$\Bigg\downarrow$ & 
$\{b_i\} = \begin{pmatrix} \frac{10}{3} \\ \frac{26}{3} \\ \frac{26}{3} \end{pmatrix}\,, \quad
\{b_{ij} \} = \begin{pmatrix}
 \frac{4447}{6} & \frac{249}{2} & \frac{249}{2} \\
 \frac{1245}{2} & \frac{779}{3} & 48 \\
 \frac{1245}{2} & 48 & \frac{779}{3} \\
\end{pmatrix}$ \\ & \\[-5mm]\hline
$G_3$ & broken at $Q=M_3$ \\\hline
& \\[-5mm]
$\Bigg\downarrow$ & 
$\{b_i\} = \begin{pmatrix}-7 \\ -2 \\ -2 \\ 7\end{pmatrix} \,, \quad
\{b_{ij} \} = \begin{pmatrix}
 -26 & \frac{9}{2} & \frac{9}{2} & \frac{1}{2} \\
 12 & 31 & 6 & \frac{27}{2} \\
 12 & 6 & 31 & \frac{27}{2} \\
 4 & \frac{81}{2} & \frac{81}{2} & \frac{115}{2}
 \end{pmatrix}$ \\[-5mm] & \\\hline
$G_2$ & broken at $Q=M_2$ \\\hline
& \\[-5mm]
$\Bigg\downarrow$ & 
$\{b_i\} = \begin{pmatrix}-7 \\ -\frac{8}{3} \\ -2 \\ \frac{11}{2}\end{pmatrix} \,, \quad
\{b_{ij} \} = \begin{pmatrix}
 -26 & \frac{9}{2} & \frac{9}{2} & \frac{1}{2} \\
 12 & \frac{37}{3} & 6 & \frac{3}{2} \\
 12 & 6 & 31 & \frac{27}{2} \\
 4 & \frac{9}{2} & \frac{81}{2} & \frac{61}{2} \\
\end{pmatrix}$ \\[-5mm] &
\\\hline
$G_1$ & broken at $Q=M_1$ \\\hline
& \\[-5mm]
$\Bigg\downarrow$ & 
$\{b_i\} = \begin{pmatrix} -7 \\ -\frac{19}{6} \\ \frac{41}{10} \end{pmatrix} \,,\quad
\{b_{ij} \} = \begin{pmatrix} 
-26 & \frac{9}{2} & \frac{11}{10} \\
12 & \frac{35}{6} & \frac{9}{10} \\
\frac{44}{5} & \frac{17}{10} & \frac{199}{50} 
\end{pmatrix}$ 
\\[-5mm] & \\\hline
$G_{\rm SM}$ & \\
\hline \hline
\end{tabular}
\end{center}
\caption{Coefficients $b_i$ and $b_{ij}$ of gauge coupling $\beta$ functions appearing in the specified breaking chain. \label{tab:beta}}
\end{table}

\begin{figure}[t!]
\centering
\includegraphics[width=.45\textwidth]{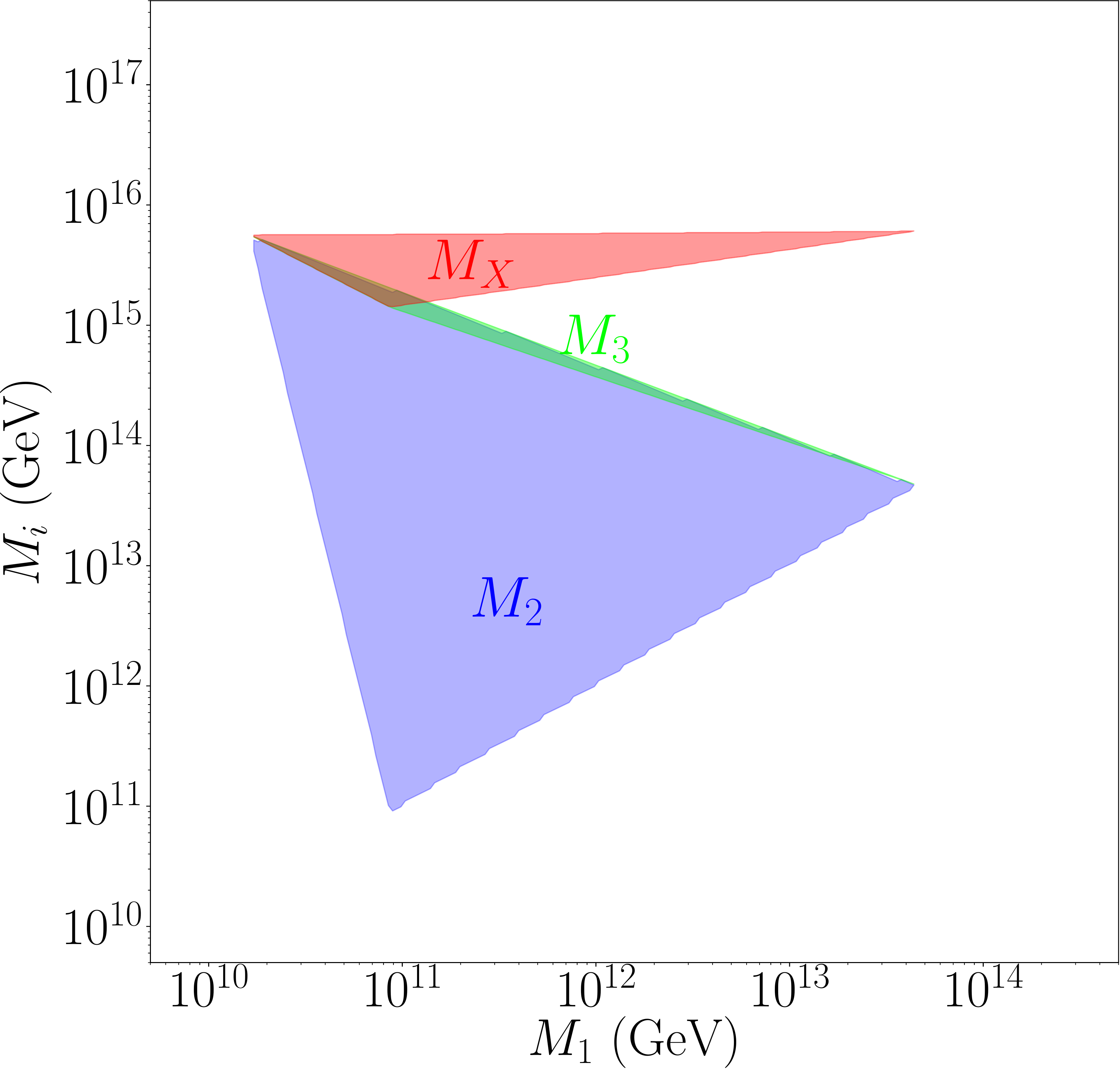}
\includegraphics[width=.45\textwidth]{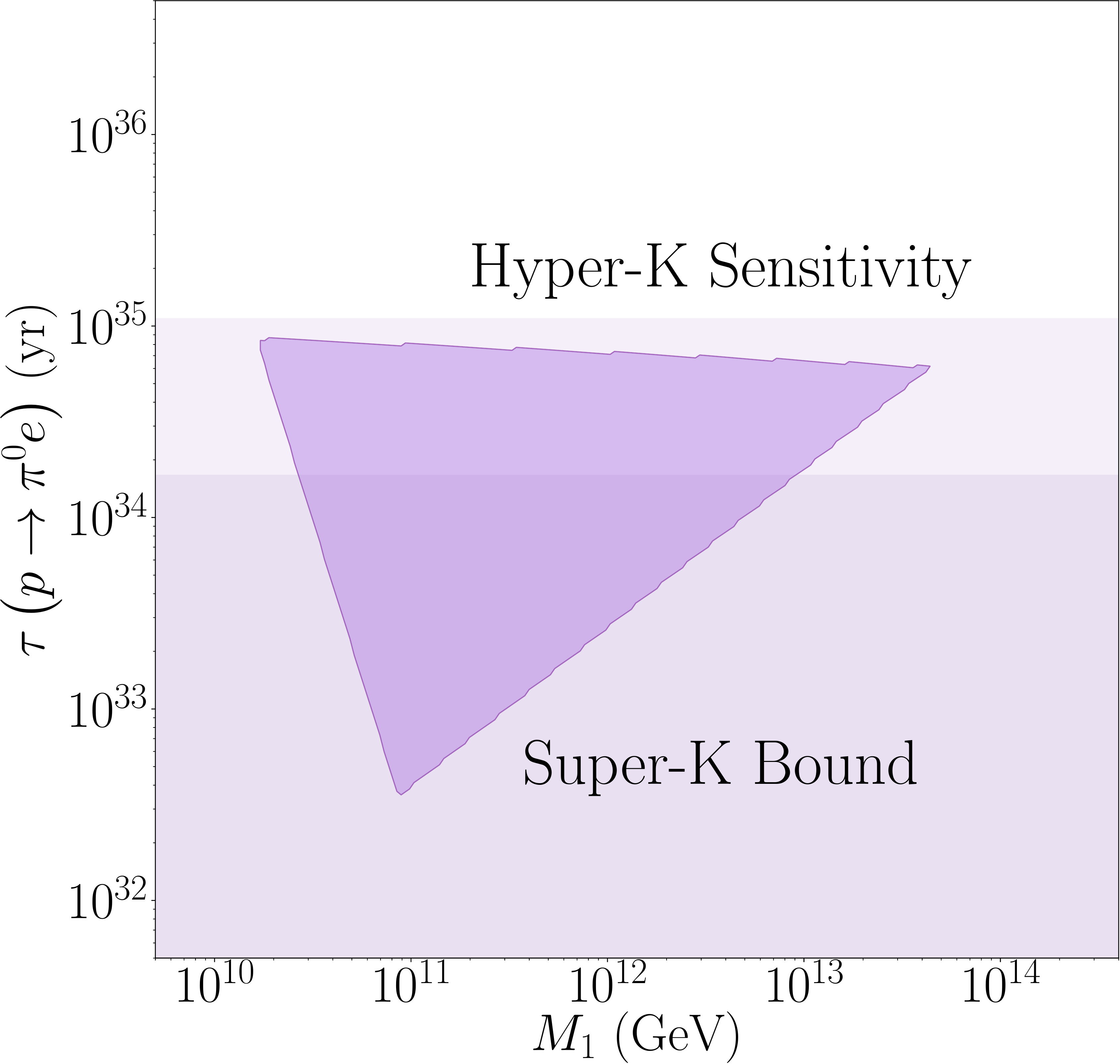}
\caption{Left panel: regions of $M_2$, $M_3$, $M_X$ as functions of $M_1$ allowed by gauge unification; Right panel: prediction of proton lifetime as functions of $M_1$, with exclusion upper bound of Super-K and future sensitivity of Hyper-K indicated.}\label{fig:Scales}
\end{figure}

During the symmetry breaking at an intermediate scale ($M_3$, $M_2$ or $M_1$), gauge couplings of the larger symmetry and those of the residual symmetry after spontaneous symmetry breaking (SSB) must satisfy matching conditions. Here we list one-loop matching conditions that appear in the GUT breaking chains. For a simple Lie group $H_{i+1}$ broken to subgroup $H_i$ at the scale $Q = M_I$, the one-loop matching condition is given by \cite{Chakrabortty:2017mgi}
\begin{eqnarray}
H_{i+1} \to H_i\,: \quad
\alpha_{H_{i+1}}^{-1}(M_I) - \frac{1}{12\pi} C_2(H_{i+1}) &=& \alpha_{H_i}^{-1}(M_I) - \frac{1}{12\pi} C_2(H_i) \,.
\end{eqnarray}
For $G_1\to G_{\rm SM}$, we encounter the breaking, $SU(2)_R \times U(1)_{X} \to U(1)_Y$, which has the matching condition \cite{Chakrabortty:2019fov}:
\begin{eqnarray}
SU(2)_R \times U(1)_X \to U(1)_Y \,: &&
\frac{3}{5} \Big(\alpha_{2R}^{-1}(M_I) -\frac{1}{6\pi} \Big) + \frac{2}{5}\alpha_{1X}^{-1}(M_I) = \alpha_{1Y}^{-1}(M_1) \,. 
\end{eqnarray}
Applying the matching conditions of the above two equations, all gauge couplings of the subgroups unify into a single gauge coupling, $\alpha_X \equiv g_X^2/{4\pi}$, of $SO(10)$ at the GUT scale, $M_X$. This condition
restricts both the GUT and intermediate scales for each breaking chain. 
We denote the mass of the heavy gauge boson masses associated with $SO(10)$ breaking as $M_X$ and $M_3$, $M_2$ and $M_1$ are associated to the breaking of $G_3$, $G_2$ and $G_1$, respectively. 
Correlations among $M_1$, $M_2$, $M_3$ and $M_X$ are determined numerically using the following procedure for the breaking chain $SO(10) \to G_3 \to G_2 \to G_1 \to G_{\rm SM}$ where the two-loop RG running evolution is performed in reverse, $G_{\rm SM} \to G_1 \to G_2 \to G_3 \to SO(10)$:
\begin{enumerate}
\item Begin the evaluation from the scale $M_Z$ with the SM gauge couplings
$\alpha_3 = 0.1184$, $\alpha_2 = 0.033819$ and $\alpha_1 = 0.010168$ \cite{Xing:2011aa}. Evolve these couplings using the RGE of the SM to scale $M_1$, where $G_1$ is recovered. Apply the matching conditions for the SM gauge couplings and the $G_1$ gauge couplings to obtain the values of couplings in the intermediate symmetry group. 
\item RG evolve the $G_1$ gauge couplings from the scale $M_1$ to $M_2$, where $G_2$
 is recovered, and the gauge couplings of $G_2$ are obtained via matching conditions at scale $M_2$. 
\item Repeating this same procedure, to evolve all couplings to the GUT scale, $M_X$, to unify to a single value $\alpha_X$ with the matching condition at $M_X$ fully accounted for.
\end{enumerate}
The above RG running procedure involves four scales $M_1$, $M_2$, $M_3$ and $M_X$. Gauge unification requires that three SM gauge couplings meet each other at the GUT scale, up to matching conditions, and enforces two constraints; thus, there are only two free scales. 
The remaining scales and the gauge coupling, $\alpha_{X}$, are then determined via gauge unification. 
General restrictions on the parameter space of scales i.e., $M_2$, $M_3$ and $M_X$ varying with $M_1$, are shown in the left plot of \figref{fig:Scales}. 

Given the gauge unification scale, $M_X$, and its gauge coupling at that scale, the proton lifetime via the decaying process $p \to \pi^0 e^+$ is predicted. Due to the scale correlations imposed by gauge unification, the correlation between the proton lifetime, $\tau_p$, and $M_X$ can be transformed into a correlation between $\tau_p$ and any intermediate scale. Following the formulation of Ref.~\cite{King:2021gmj}, we derive the allowed parameter space of $\tau_p$ versus intermediate scales. 
By varying the lowest intermediate scale $M_1$, we obtain general regions of $\tau_p$. The bound of $\tau_p$ versus the lowest scale $M_1$ is shown in the right panel of \figref{fig:Scales}. 
The Super-K experiment set a lower bound on the proton lifetime, $\tau(p\to e^+ \pi^0) > 2.4 \times 10^{34}$~years at 90 \% confidence level \cite{Super-Kamiokande:2020wjk}. In the future, Hyper-Kamiokande (Hyper-K) is expected to improve the measurement of proton lifetime by almost one order of magnitude \cite{Hyper-Kamiokande:2018ofw}. 
If proton decay is not observed, the entire parameter space of this breaking chain will be excluded.

\begin{figure}[t!]
\centering
\includegraphics[width=.7\textwidth]{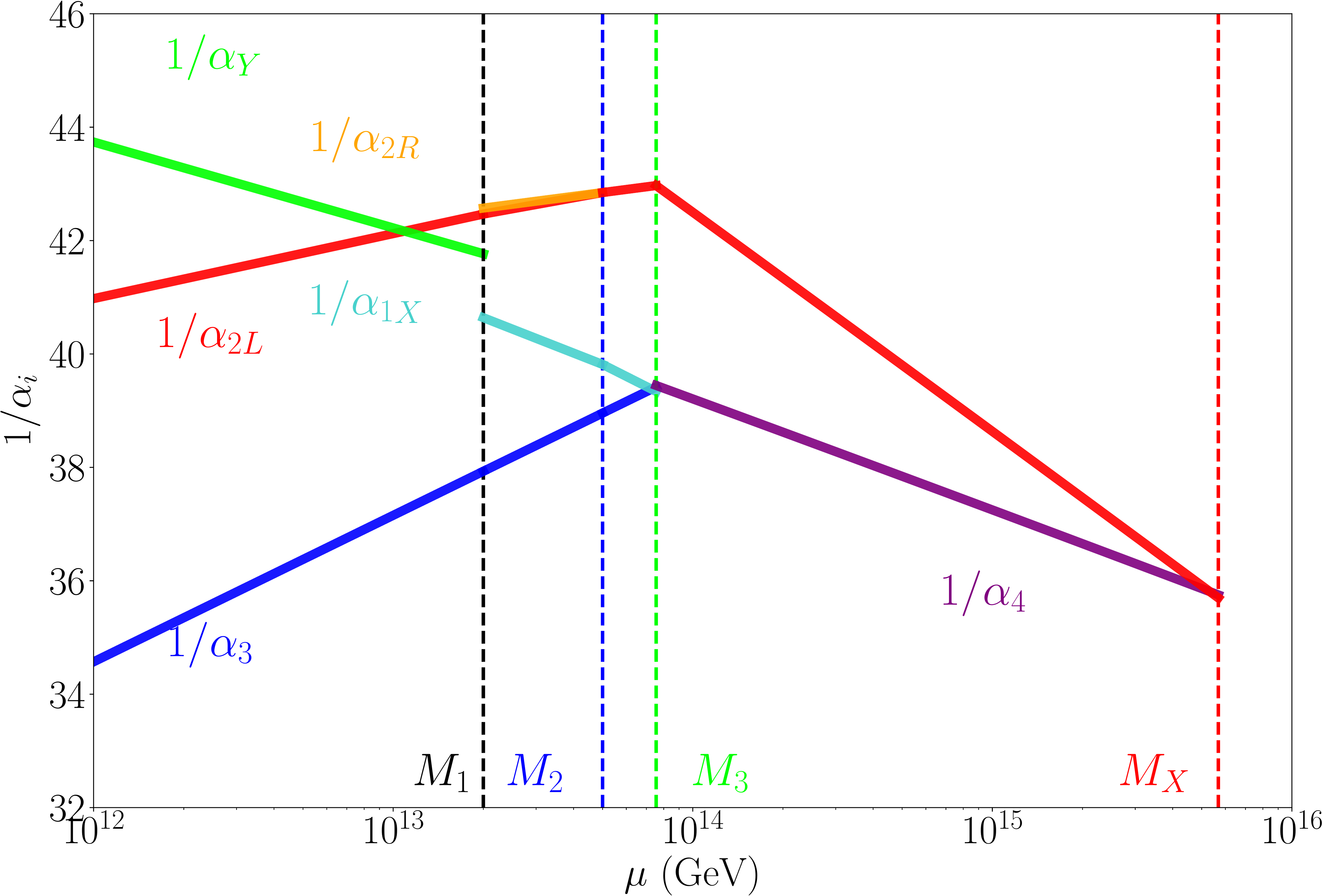}
\caption{The RG running of gauge couplings in the breaking chain $SO(10) \to G_3 \to G_2 \to G_1 \to G_{\rm SM}$. BP1 with the first and second lowest intermediate scales are fixed at $M_1 = 2 \times 10^{13}$~GeV and $M_2 = 5 \times 10^{13}$~GeV, the remaining scales $M_3$ and $M_X$, as well as gauge couplings $\alpha_{2R}$, are determined by the gauge unification at $M_X$.}\label{fig:RGE_2}
\end{figure}

{\bf Benchmark Point 1 (BP1)} In \figref{fig:RGE_2}, we show an example of the RG running of the gauge couplings along with the scale and fix
\begin{eqnarray}
M_1 = 2 \times 10^{13}~{\rm GeV} \,,\quad 
M_2 = 5 \times 10^{13}~{\rm GeV}\,,
\end{eqnarray} where the remaining scales, $M_3$ and $M_X$, as well as the gauge coupling $\alpha_{X}$, are then determined via the gauge unification,
\begin{eqnarray}
M_3 = 7.55 \times 10^{13}~{\rm GeV} \,,\quad 
M_X = 5.68 \times 10^{15}~{\rm GeV} \,,\quad 
\alpha_X = 0.0279\,.
\end{eqnarray} 
This benchmark point will be considered throughout this paper. Its associated proton decay rate, 
$\tau(p\to e^+ \pi^0) \sim 5.1\times 10^{34}$ years, is consistent with the current Super-K bound and will be tested by Hyper-K. We note that BP1 is consistent with SM fermion masses and mixing to a high statistical significance, and this requires $M_{1}\sim 10^{13}$ GeV. Such a high value for $M_1$ leads a compressed hierarchy between $M_{1}$, $M_{2}$ and $M_{3}$ and this comes from the constraint of gauge unification (from the left panel of \figref{fig:Scales} this region is in the right corner of the blue triangle.)

\section{Fermion masses and mixing}\label{sec:fandf}
As all the SM fermions are embedded in the same $SO(10)$ multiplet (${\bf 16}$), their masses are correlated with each other. Therefore, it is a non-trivial task to find regions of the GUT model parameter space that predict the SM fermion masses and mixing consistent with the precisely measured (particularly in the quark sector) experimental data. This section presents the correlations of masses and mixing between quarks and leptons and predicts heavy neutrino masses using the model we discussed in the previous section. 
We parametrise the up, down, neutrino, charged lepton Yukawa couplings and right-handed neutrino mass matrix, respectively, as follows \cite{Altarelli:2010at}: 
\be
\begin{alignedat}{3}\label{eq:Yukawa_Hermitian}
& Y_u = h + r_2 f + i\, r_3 h^\prime \,, \quad
&& Y_d = r_1 (h+ f + i\, h^\prime) \,, \quad
&&Y_\nu = h - 3 r_2 f + i\, c_\nu h^\prime\,, \\
& Y_e = r_1 (h - 3 f + i\, c_e h^\prime) \,, \quad
&& M_{\nu_R} = f \, \frac{\sqrt{3} \,r_1}{V_{16}} v_S\,,
\end{alignedat}  
\ee
where 
\be
\begin{alignedat}{4}
& h = Y_{\bf 10} V_{11} \,, \;
&& f = Y_{\overline{\bf 126}} \frac{V_{16}}{\sqrt{3}} \frac{V_{11}^*}{V_{15}} \,,\;
&& c_e= \frac{V_{17}- \sqrt{3} V_{18}}{V_{17}+ V_{18}/\sqrt{3}}\,, \quad
&&c_\nu= \frac{V_{13}^*- \sqrt{3} V_{14}^*}{V_{17}+ V_{18}/\sqrt{3}} \frac{V_{15}}{V_{11}^*}\,,\\
& r_1 = \frac{V_{15}}{V_{11}^*} \,, \quad
&& r_2 = \frac{V_{12}^*}{V_{16}} \frac{V_{15}}{V_{11}^*}\,,\quad
&& r_3 = \frac{V_{13}^*+ V_{14}^*/\sqrt{3}}{V_{17}+ V_{18}/\sqrt{3}} \frac{V_{15}}{V_{11}^*}\,, \;
&& h^\prime= -i\, Y_{\bf 120} \left(V_{17}+ V_{18}/\sqrt{3}\right) \frac{V_{11}^*}{V_{15}}\,, \quad
\end{alignedat} 
\ee
and $V_{ji}$ denotes the mixing between the mass and interaction basis of the electroweak Higgs doublets. The light neutrino mass matrix, $M_\nu$, is obtained by
\begin{eqnarray}
M_\nu &=&
m_0 Y_\nu f^{-1} Y_\nu\,,
\end{eqnarray} 
where $m_0= - \frac{V_{16}}{\sqrt{3} \,r_1} \frac{v_{\rm SM}^2}{v_S}$.

\subsection{Parametrisation using Hermitian Yukawa matrices}

The most general form of Yukawa couplings and neutrino mass matrix includes many free parameters.
A considerable reduction in the number of parameters can be achieved by considering only
the Hermitian case for all fermion Yukawa couplings matrices $Y_u$, $Y_d$, $Y_\nu$ and $Y_e$ (and $M_R$ should be real as a consequence of the Majorana nature for right-handed neutrinos).
Such a reduction can result from spontaneous CP violation \cite{Grimus:2006bb,Grimus:2006rk} which assumes that there exists a CP symmetry above the GUT scale, leading to real-valued $Y_{\bf 10}$, $Y_{\overline{\bf 126}}$ and $Y_{\bf 120}$, and the CP is broken by some complex VEVs of Higgs multiplets during GUT or intermediate symmetry breaking.
For the particular chain we applied in the last section, one can consider, for example, the parity-odd singlet of $G_2 \equiv SU(3)_c \times SU(2)_L \times SU(2)_R \times U(1)_X \times Z_2^C$, decomposed from ${\bf 45}$, gains a purely imaginary VEV. Then, via couplings such as ${\bf 45} \cdot {\bf 10} \cdot {\bf 120}$ (and ${\bf 45} \cdot \overline{\bf 126} \cdot {\bf 120}$) which generate purely imaginary off-diagonal mass terms between $h_{\bf 10}^{u,d}$ and $h_{\bf 120}^{u,d}$ (and those between $h_{\overline{\bf 126}}^{u,d}$ and $h_{\bf 120}^{u,d}$) and further purely imaginary mixing entries $V_{13}$, $V_{14}$ (and $V_{17}$, $V_{18}$) are obtained.
As a result, $h$, $f$ and $h'$, as well as all parameters on the right-hand side of Eq.~\eqref{eq:Yukawa_Hermitian}, are real. Since $h'$ is antisymmetric, we arrive at Hermitian Dirac Yukawa coupling matrices $Y_u$, $Y_d$, $Y_\nu$ and $Y_e$. This texture has widely been applied in the literature, see e.g., Refs.~\cite{Dutta:2004hp,Dutta:2004zh,Joshipura:2011nn}. The resulting fermion mass matrices conserve parity symmetry $L \leftrightarrow R$ \cite{Dutta:2004hp} and following from the assumption that there is no CP violation in the Higgs sector, apart from that of ${\bf 120}$, $r_1$, $r_2$, $r_3$, $c_e$, and $c_\nu$ are all real parameters
resulting in a real symmetric right-handed neutrino mass matrix, $M_{\nu_R}$. 
The CP symmetry in the Yukawa coupling is spontaneously broken after the Higgses gain VEVs. 

For simplicity, we assume that $r_3 = 0$, which implies that the imaginary part of $Y_u$ vanishes. It is convenient to write the up-type Yukawa in the diagonal basis
\begin{eqnarray}
Y_u = h + r_2 f = {\rm diag}\{ \eta_u y_u, \eta_c y_c, \eta_t y_t\}\,,
\end{eqnarray}
which can be achieved via a real-orthogonal transformation on the fermion flavours without changing the Hermitian property of $Y_d$, $Y_e$, and $Y_\nu$.
In the above, $\eta_{u,c,t} = \pm 1$ refer to signs that cannot be determined by the real-orthogonal transformation. While $\eta_t = +1$ can be fixed by making an overall sign rotation for all Yukawa matrices, the remaining signs, $\eta_u$ and $\eta_c$, cannot be fixed and are randomly varied throughout our analysis.
In the basis of the diagonal up-quark mass matrix, $Y_d$ is given by
\begin{eqnarray}
Y_d = P_a V_{\rm CKM} \, {\rm diag}\{ \eta_d y_d, \eta_s y_s, \eta_b y_b\} \,V_{\rm CKM}^\dag P_a^* \,,
\end{eqnarray} 
where again $\eta_{d,s,b} = \pm 1$ represent the signs of eigenvalues, and $V_{\rm CKM}$ is the CKM matrix parametrised in the following form
\begin{eqnarray}
V_{\rm CKM} = \left(
\begin{array}{ccc}
 c_{12} c_{13} & s_{12} c_{13} & s_{13} e^{-i \delta_q } \\
 - s_{12} c_{23}- c_{12} s_{13} s_{23} e^{i \delta_q } & c_{12} c_{23}- s_{12} s_{13} s_{23} e^{i \delta_q } & c_{13} s_{23} \\
 s_{12} s_{23}- c_{12} s_{13} c_{23} e^{i \delta_q } & - c_{12} s_{23}- s_{12} s_{13} c_{23} e^{i \delta_q } & c_{13} c_{23} \\
\end{array}
\right) \,,
\end{eqnarray} 
where $s_{ij} = \sin \theta_{ij}^q$, $c_{ij} = \cos \theta_{ij}^q$ and 
$P_a= {\rm diag}\{ e^{ia_1}, e^{ia_2}, 1\}$. The matrices $h$, $f$ and $h^\prime$ are then expressed in terms of $Y_u$ and $Y_d$
\begin{alignat*}{3}
& h =- \frac{Y_u}{r_2-1} + \frac{r_2 {\rm Re} Y_d}{r_1(r_2-1)} \,, \quad 
&& f =\frac{Y_u}{r_2-1}- \frac{{\rm Re} Y_d}{r_1(r_2-1)} \,, \quad 
&& h^\prime = i \, \frac{{\rm Im} Y_d}{r_1}\,,
\end{alignat*}
where $Y_\nu$, $Y_e$ are 
\begin{eqnarray} \label{eq:Yukawa_lepton}
Y_\nu &=& - \frac{3r_2+1}{r_2-1} Y_u + \frac{4r_2}{r_1(r_2-1)}{\rm Re} Y_d + i \frac{c_\nu}{r_1} {\rm Im} Y_d \,, \nonumber\\
Y_e &=& - \frac{4r_1}{r_2-1} Y_u + \frac{r_2+3}{r_2-1}{\rm Re} Y_d +i c_e {\rm Im} Y_d\,.
\end{eqnarray} 
The light neutrino mass matrix can be expressed as
\begin{eqnarray} \label{eq:M_nu}
M_\nu &=& m_0 \left( \frac{8r_2(r_2+1)}{r_2-1}Y_u - \frac{16 r_2^2}{r_1(r_2-1)} {\rm Re} Y_d \right. \nonumber\\
&& \left.+ \frac{r_2-1}{r_1} \left(r_1Y_u + i c_\nu {\rm Im} Y_d \right) \left(r_1Y_u - {\rm Re} Y_d \right)^{-1} \left(r_1Y_u - i c_\nu {\rm Im} Y_d \right) \right)\,.
\end{eqnarray} 
Using this parametrisation, all six quark masses and four CKM mixing parameters are treated as inputs, and we are then left with seven parameters ($a_1$, $a_2$, $r_1$, $r_2$, $c_e$, $c_\nu$, and $m_0$) to fit eight observables, including three Yukawa couplings $y_e$, $y_\mu$, $y_\tau$, two neutrino mass-squared differences $\Delta m^2_{21}$, $\Delta m^2_{31}$ and three mixing angles $\theta_{12}$, $\theta_{13}$, $\theta_{23}$, where the leptonic CP-violating phase, $\delta$, will be treated as a prediction\footnote{While we do not show the Majorana phases, we compute the effective Majorana mass.}.

\subsection{Procedure of numerical analysis}
\begin{figure}[h!]
\centering
\includegraphics[width=.85\textwidth]{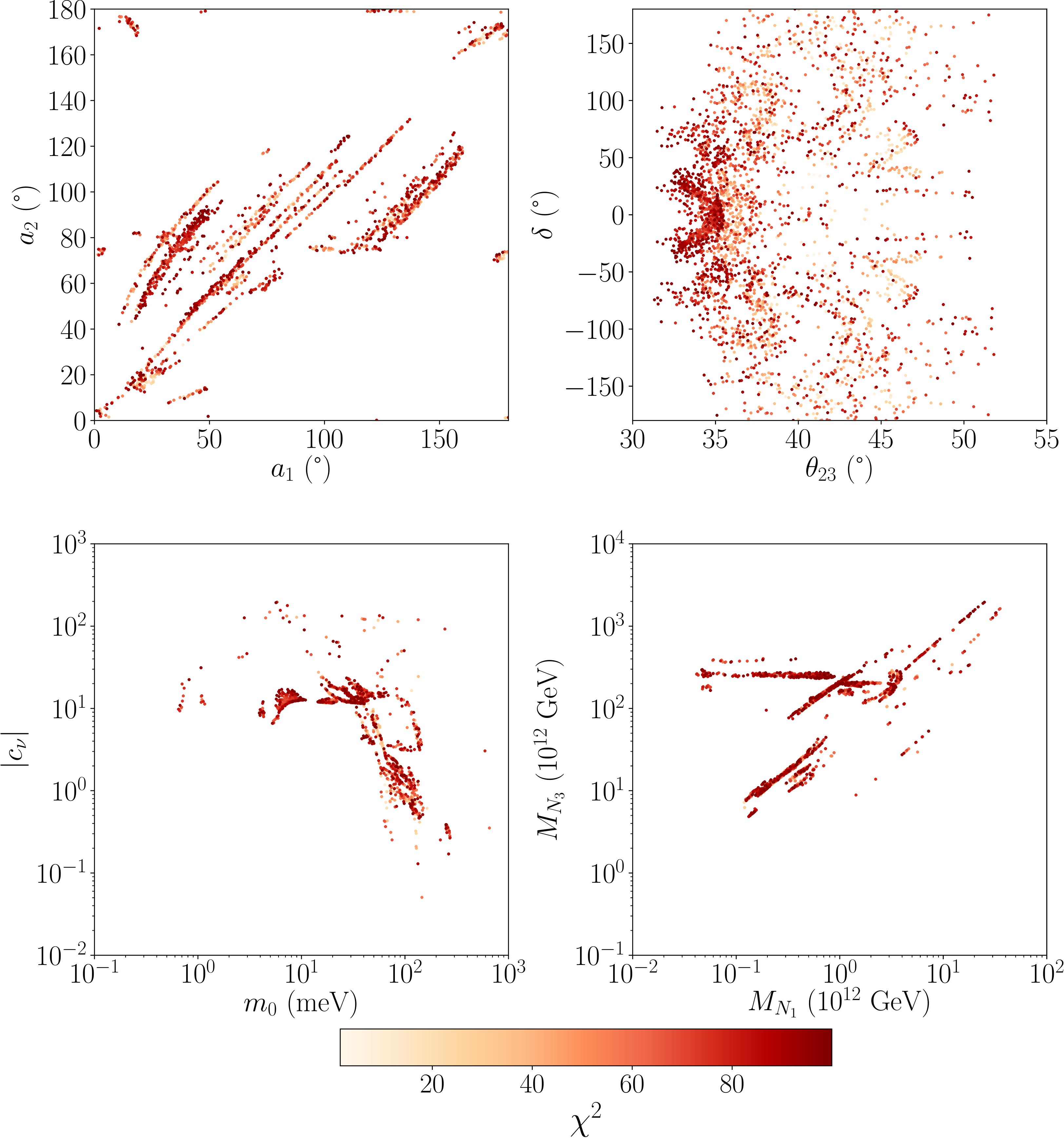}
\caption{Two-dimensional correlations between theory inputs (left two panels) and predicted observables (right two panels) for $\chi^2<100$ for $\theta_{23}\leq 45^{\circ}$. Consistency with gauge unification is not considered.}\label{fig:chisq_100}
\end{figure}

This section describes how we identify regions of our model parameter space consistent with fermion masses and mixing while evading the existing proton decay limit. In our numerical analysis, we use the following experimental data:
\begin{itemize}
\item
We fix the Yukawa couplings ($y$) of charged fermions and CKM mixing angles ($\theta$) at their best-fit (bf) values \cite{Xing:2007fb,Joshipura:2011nn,Babu:2016bmy}
\be
\begin{alignedat}{3} \label{eq:Yukawa_values}
& y_u^{\rm bf} = 2.54 \times 10^{-6} \,, \quad
&& y_c^{\rm bf} = 1.37 \times 10^{-3} \,, \quad
&& y_t^{\rm bf} = 0.43\,, \\
& y_d^{\rm bf} = 6.56 \times10^{-6} \,, \quad
&& y_s^{\rm bf} = 1.24 \times10^{-4} \,, \quad
&& y_b^{\rm bf} =5.7 \times 10^{-3}\,, \\
& y_e^{\rm bf} = 2.70 \times 10^{-6} \,, \quad
&& y_\mu^{\rm bf} = 5.71 \times 10^{-4}\,, \quad
&& y_\tau^{\rm bf} = 9.7\times 10^{-3}\,, \\
\end{alignedat} 
\ee
and
\begin{alignat}{4} 
& \theta^{q, {\rm bf}}_{12} = 0.227 \,, \quad
&& \theta^{q, {\rm bf}}_{23} = 4.858\times10^{-2} \,, \quad
&& \theta^{q, {\rm bf}}_{13} = 4.202\times10^{-3}\,, \quad
&& \delta^{q, {\rm bf}}=1.207 \,. \quad
\end{alignat} 
These values are obtained by RG evolving the experimental best-fit values at a low scale to $2\times 10^{16}$~GeV, where we have ignored the experimental errors. For simplicity, small corrections induced by RG running above intermediate scales have been ignored, but their inclusion would further relax the parameter space.\footnote{Although the coupling for the heaviest RH neutrino in Eq.(2.6) can be of order 1, its contribution to RG running is under control and is expected to be at most  5\% as $M_X/M_1\sim10^{-2}$.} However, as we will later see, fixing them at the best fit values is sufficient to reproduce all mixing data. Thus, in this current discussion, we will ignore them for simplicity. 

\item In the neutrino sector, we use the best-fit values from NuFIT 5.1 \cite{Esteban:2020cvm} and include the $1\sigma$ uncertainty. Those data with and without Super-K atmospheric data are, respectively, given by
\begin{eqnarray} \label{eq:lepton_data_1}
&\Delta m^2_{21} = (7.42 \pm 0.21) \times 10^{-5}~ {\rm eV}^2\,, \;
\Delta m^2_{3l} = (2.510 \pm 0.027) \times 10^{-3}~ {\rm eV}^2 \,, \nonumber\\
&\theta_{12} = 33.45^\circ \pm 0.77^\circ\,, \qquad
\theta_{23} = 42.1^\circ \pm 1.1^\circ\,, \qquad
\theta_{13} = 8.62^\circ \pm 0.12^\circ\,,
\end{eqnarray}
and 
\begin{eqnarray} \label{eq:lepton_data_2}
&\Delta m^2_{21} = (7.42 \pm 0.21) \times 10^{-5}~ {\rm eV}^2\,, \;
\Delta m^2_{3l} = (2.514 \pm 0.028) \times 10^{-3}~ {\rm eV}^2 \,, \nonumber\\
&\theta_{12} = 33.44^\circ \pm 0.77^\circ\,, \qquad
\theta_{23} = 49.0^\circ \pm 1.3^\circ\,, \qquad
\theta_{13} = 8.57^\circ \pm 0.13^\circ\,,
\end{eqnarray}
The atmospheric mixing angle, $\theta_{23}$, is restricted to first octant ($0 <\theta_{23} < 45^\circ$) and the second ($45^\circ <\theta_{23} < 90^\circ$), respectively, in the two cases.  
In both cases, normal ordering (i.e., $m_1 < m_2 < m_3$) of neutrino masses is assumed. Inverted ordering (i.e., $m_3 < m_1 < m_2$) will not be discussed as a preliminary scan indicates that our model does not favour the inverted ordering. We do not consider the small flavour-dependent RG running effect due to the suppression of charged lepton Yukawa coupling. 
 \end{itemize}
The statistical analysis is performed in the following way:
\begin{itemize}
\item 
As quark masses and mixing parameters are fixed at their best-fit values, $Y_u$ is fully determined except for the signs of $\eta_u$ and $\eta_d$ (note that $\eta_t = +1$ is fixed by an overall sign rotation).
$Y_d$ depends on two free model parameters, $a_1$ and $a_2$, and signs $(\eta_d,\eta_s,\eta_b)$. 
\item
Based on Eq.~\eqref{eq:Yukawa_lepton}, $Y_e$ depends on the two phases $a_1$, $a_2$ and three ratios $r_1$, $r_2$, $c_e$ up to the above sign differences. Note that $Y_e$ must satisfy 
 three equations simultaneously:
\begin{eqnarray}\label{eq:charged_lepton_yukawas}
 {\rm Tr}[Y_e Y_e^\dag] &=& y_e^2 + y_\mu^2 + y_\tau^2 \,, \nonumber\\
 {\rm Tr}[Y_e Y_e^\dag Y_e Y_e^\dag] &=& y_e^4 + y_\mu^4 + y_\tau^4 \,,\nonumber\\
 {\rm Det}[Y_e Y_e^\dag] &=& y_e^2 y_\mu^2 y_\tau^2 \,,
\end{eqnarray}
 and as the right hand side is fixed, $r_1$, $r_2$ and $c_e$ are fully determined by the phases $a_1$, $a_2$ and the signs $\eta_q$ (for $q= u,c,d,s,b$). We scan the phase parameters in the range $a_1, a_2 \in [ 0, 2\pi ]$ and vary the signs $\eta_q = \pm 1$ randomly and solve for $r_1$, $r_2$ and $c_e$. Then, we substitute these values into Eq.~\eqref{eq:Yukawa_lepton} and determine the unitary matrix $V_e$ used in the diagonalisation $V_e^\dag Y_e Y_e^\dag V_e = {\rm diag} \{ y_e^2, y_\mu^2, y_\tau^2 \}$. 

\item In Eq.~\eqref{eq:M_nu}, the neutrino mass matrix, $M_\nu$, is determined by two further parameters $c_\nu$ and $m_0$. The former determines the flavour structure and the latter the absolute mass scale, and by scanning these parameters, we determine $M_\nu$. The diagonalisation $V_\nu^\dag M_\nu V_\nu^* = {\rm diag} \{ m_1, m_2, m_3 \}$ provides the neutrino mass eigenvalues and unitary matrix $V_\nu$. 

\item The PMNS matrix is given by $U_{\rm PMNS} = V_e^\dag V_\nu$, and the three leptonic mixing angles are derived via 
\begin{eqnarray}
\sin \theta_{13} = |(U_{\rm PMNS})_{e3}|\,,~
\tan \theta_{12} = \left|\frac{(U_{\rm PMNS})_{e2}}{(U_{\rm PMNS})_{e1}} \right| \,,~
\tan \theta_{23} = \left|\frac{(U_{\rm PMNS})_{\mu 3}}{(U_{\rm PMNS})_{\tau 3}} \right| \,. 
\end{eqnarray}
These angles and two mass squared differences $\Delta m^2_{21} = m^2_2- m_1^2$ and $\Delta m^2_{31} = m^2_3- m_1^2$ are taken as outputs to compare with the experimental data shown in \equaref{eq:lepton_data_1}. 
\end{itemize}
In summary, once the charged fermion masses and quark mixing parameters are fixed, we are left with only four free model parameters $a_1, a_2, c_\nu, m_0$ and signs $\eta_q$:
\begin{eqnarray}
{\cal P}_m \in \{a_1, a_2, c_\nu, m_0, \eta_q\}\,.
\end{eqnarray} 
We scan the model parameter space, ${\cal P}_m$, to fit five observables: 
\begin{eqnarray}
{\cal O}_n \in \{\theta_{12}, \theta_{13}, \theta_{23}, \Delta m^2_{21}, \Delta m^2_{31}\}\,.
\end{eqnarray}
In this way, we efficiently reduce the dimensionality of the parameter space from 17 to 5 dimensions.
Following the above simplified treatment, we scan two phases $a_1, a_2$ in the range $[0, \pi]$.
The coefficient $|c_\nu|$ is logarithmically scanned in the range [$10^{-3}\,, 10^3$], and we randomly assign its $\pm$ sign. $m_0$ (meV) is solved by minimising the $\chi^2$ function, which is used as a measure of how well our model fits the data, being defined as
\begin{eqnarray} 
\chi^2 = \sum_n \left[ \frac{{\cal O}_n({\cal P}_m)-{\cal O}_n^{\rm bf}}{\sigma_{{\cal O}_n}} \right]^2 \,.
\end{eqnarray} 
Given the predefined theory model parameter space, ${\cal P}_m$, and scanning in the relevant ranges of these parameters, 
we determine which regions fit the experimental data by setting an upper bound of $\chi^2$ value.
This procedure of the scan is divided into two steps:
We first perform a preliminary scan by setting the upper bound of $\chi^2 <100$ and then perform a subsequent scan to find the points with $\chi^2 <10$. The results of the first scan which uses the neutrino oscillation data of Eq.~\eqref{eq:lepton_data_1} (first octant) are shown in Fig.~\ref{fig:chisq_100}. A two-dimensional subspace of $a_1$-$a_2$ ($m_0$-$c_\nu$) is shown in the top (bottom) left panel and predictions of $\theta_{23}$-$\delta$ ($M_{N_1}$-$M_{N_3}$) are given in the right top (bottom) panel. 
$M_{N_1}$, $M_{N_2}$ and $M_{N_3}$ are three right-handed neutrino masses ordered from lightest to heaviest, and they are obtained by solving the inverse of the Type-I seesaw formula: 
\be
M_{\nu_R} = Y_\nu^T M_\nu^{-1} Y_\nu v_{\rm SM}^2\,,
\ee
where the mass states of $\nu_R$, from the lightest to heaviest, are denoted as $N_1$, $N_2$ and $N_3$. 
 We impose an upper bound by requiring $M_{N_3} \lesssim M_1$, and this is
approximately equivalent to requiring that the largest eigenvalue of $Y_{\overline{\bf 126}}\lesssim 1$ such that the perturbativity is respected. Since the maximal value of $M_1$ allowed by proton decay measurements is given by $4.4 \times 10^{13}~{\rm GeV}$ \cite{King:2021gmj}, viable points in the model parameter space require that
\begin{eqnarray} \label{eq:RHN_bound}
M_{N_3} < 4.4 \times 10^{13}~{\rm GeV} \,.
\end{eqnarray}
Naively, by assuming the magnitude of the Dirac Yukawa coupling $Y_\nu \sim {\cal O}(1)$, we know from the seesaw formula that the RHN mass scale is around $10^{15}$~GeV. Thus, one can expect that the condition of Eq.~\eqref{eq:RHN_bound} rules out most points.  
This is confirmed by the bottom-left panel \figref{fig:chisq_100} where most of the points predict the heaviest neutrino mass, $M_{N_3}$, to be heavier than $4.4 \times 10^{13}$~GeV. Therefore, these points are not consistent with the requirement of gauge unification. 
\begin{figure}[ht]
\centering
\includegraphics[width=.99\textwidth]{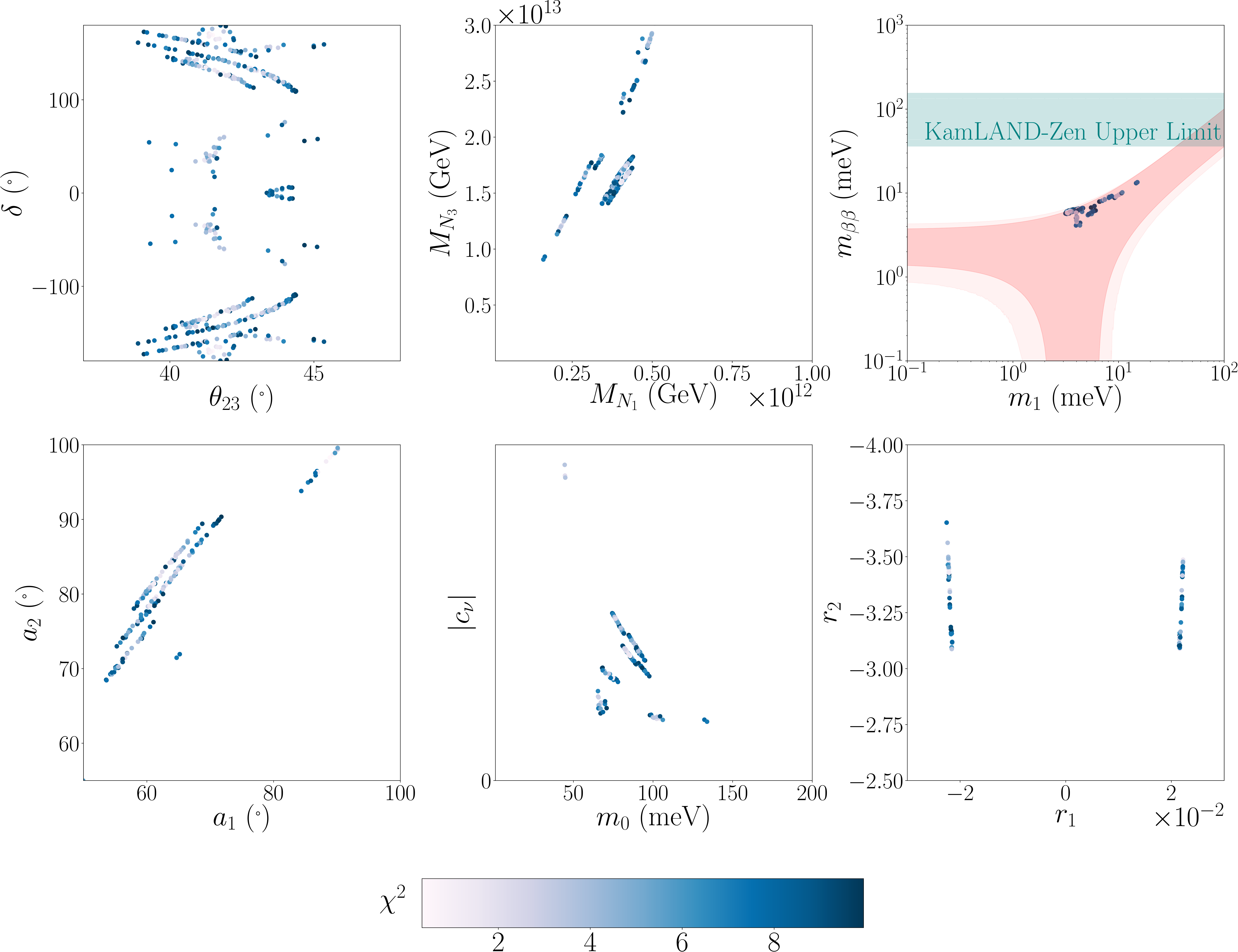}
\caption{The predicted observables (top left two panels), the effective neutrino mass prediction (top right panel) and two-dimensional correlations between theory inputs (bottom panels) for $\chi^2<10$ and $\theta_{23}\leq 45^{\circ}$. Consistency with gauge unification is considered.}
\label{fig:chisq_10_octant1}
\end{figure}
\begin{figure}[ht]
\centering
\includegraphics[width=.99\textwidth]{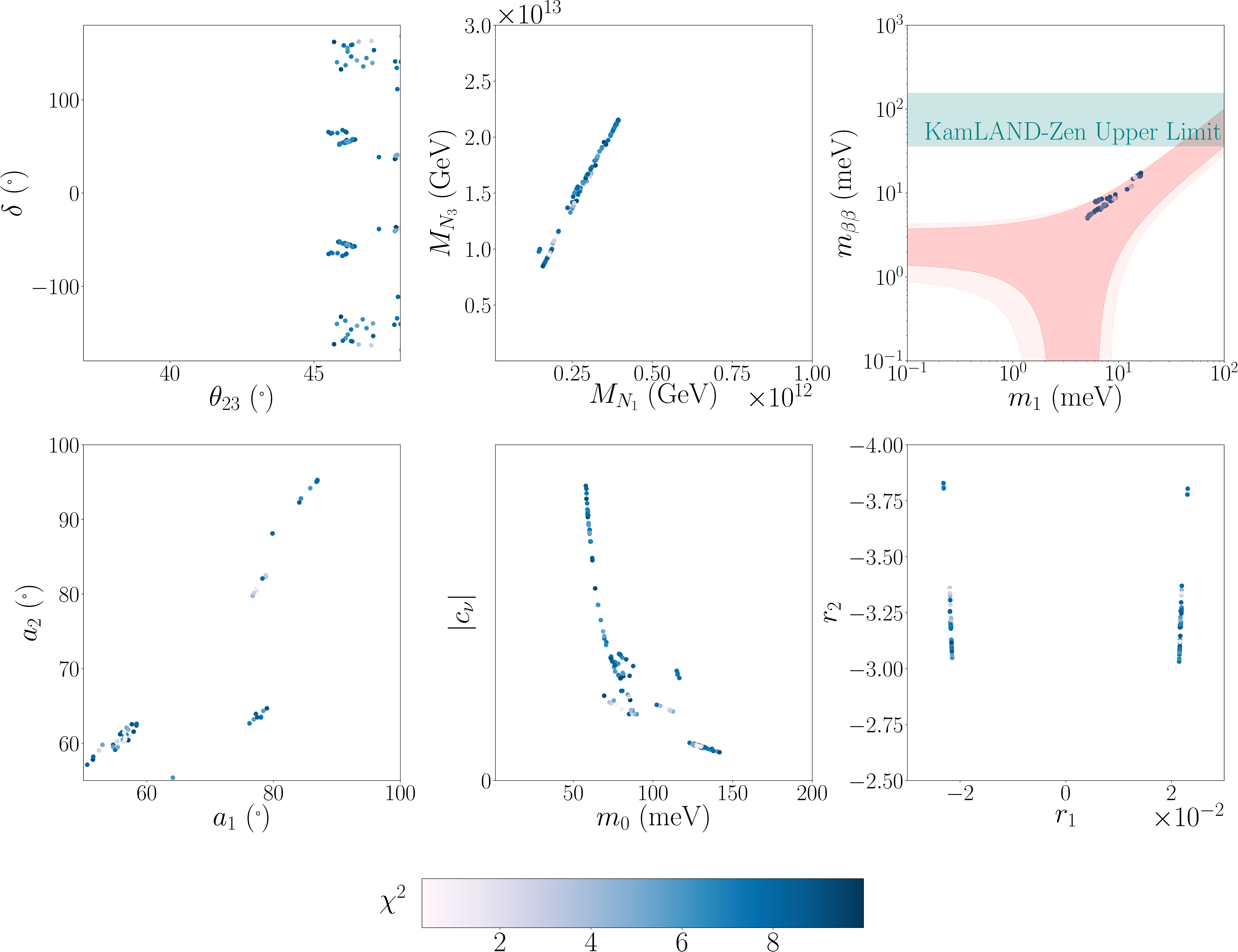}
\caption{The predicted observables (top left two panels), the effective neutrino mass prediction (top right panel) and two-dimensional correlations between theory inputs (bottom panels) for $\chi^2<10$ and $\theta_{23}\geq 45^{\circ}$. Consistency with gauge unification is considered.}
\label{fig:chisq_10_octant2}
\end{figure}
We then perform a second more dense scan around the former points by requiring $\chi^2 <10$ and gauge unification, e.g., the bound of the heaviest right-handed neutrino mass satisfying Eq.~\eqref{eq:RHN_bound}. The results of this scan are shown in Figs.~\ref{fig:chisq_10_octant1} and \ref{fig:chisq_10_octant2}, where neutrino oscillation data in Eqs.~\eqref{eq:lepton_data_1} and \eqref{eq:lepton_data_2} are used, respectively. In both figures, scatter plots of parameters are shown in the left panel and predictions of observables are given in the right panel. In the first $2 \times 2$ grid of both figures, we arrange two-dimensional subspaces of $a_1$-$a_2$, $m_0$-$c_\nu$ (left), and predictions $\theta_{23}$-$\delta$, $M_{N_1}$-$M_{N_3}$ (right).
We have checked that truncating the upper bound of $\chi^2$ from 100 to 10 removes most of the points allowed in the preliminary scan. In particular, this truncation automatically removes most points which do not satisfy the gauge unification requirement, i.e., Eq.~\eqref{eq:RHN_bound}. While only a few points are left with a large value of $M_{N_3}$, the gauge unification requirement fully removes them. Only a single large island of points and a few small islands nearby in the parameter space of $M_{N_1}$-$M_{N_3}$, where $M_{N_3}$ are predicted in the range $(1, 2)\times 10^{13}$~GeV. By comparing \figref{fig:chisq_10_octant2} with \figref{fig:chisq_100}, all the free parameters are strongly restricted, in particular, $a_1, a_2 \sim (50^\circ, 100^\circ)$, and $|c_\nu|\sim (1,10)$. Notably, these points are consistent with fermion masses and mixing, gauge unification and the experimental result of proton decay measurements.
In the bottom left panel of ~\figref{fig:chisq_10_octant2} we observe a
linear-correlation between $r_1$ and $r_2$ which we analytically derive as follows. 
From experimental quark data, the following hierarchical relations exist among mixing parameters:
\begin{eqnarray} \label{eq:hierarchy}
y_u: y_c: y_t &\sim& \theta_C^8: \theta_C^4: \theta_C^0 \,,\nonumber\\
y_d: y_s: y_b &\sim& \theta_C^8: \theta_C^6: \theta_C^3 \,,\nonumber\\
\theta_{13}^q: \theta_{23}^q: \theta_{12}^q &\sim& \theta_C^3: \theta_C^2: \theta_C^1 \,,
\end{eqnarray}
where $\theta_C$ is the Cabibbo angle. Taking these relations into account, we derive
\begin{eqnarray}
\frac{\tilde{y}_\tau}{\tilde{y}_b} \approx \frac{r_2+3}{r_2-1} \, p - \frac{4 r_1}{r_2 - 1} \frac{\tilde{y}_t}{\tilde{y}_b} \,,
\end{eqnarray}
and from Eq.~\eqref{eq:Yukawa_lepton}, where $\tilde{y}_f = \eta_f y_f $ where $\eta_f = \pm 1$ and
\begin{eqnarray}
p = 1-\frac{\tilde{y}_\mu}{\tilde{y}_\tau} + \frac{r_2+3}{r_2-1} \, \frac{\tilde{y}_s}{\tilde{y}_\tau}\,,
\end{eqnarray}
which parametrises small deviations from 1 due to the suppressions of $y_\mu / y_\tau$ and $y_s / y_\tau$. We keep this deviation due to the fact that $y_\tau/y_b \approx 1$. 
We rewrite it in the form
\begin{eqnarray}
r_2 \approx \frac{-4 \tilde{y}_t}{\tilde{y}_\tau - \tilde{y}_b p} r_1 + \frac{\tilde{y}_\tau + 3\tilde{y}_b p}{\tilde{y}_\tau - \tilde{y}_b p} \,,
\end{eqnarray}
which provides an almost linear correlation between $r_1$ and $r_2$, which is clearly shown in \figref{fig:chisq_10_octant2} for both $|\tilde{y}_\tau - \tilde{y}_b| = y_\tau - y_b$ and $y_\tau + y_b$. 
To satisfy the experimental constraint that $y_\tau/y_b\gtrsim 1$, most of the points with low $\chi^2$ values satisfy the following hierarchical relation:
\be\label{eq:r_1r_2}
r_1 \ll 1 \ll r_2 \,,
\ee
namely that ${r_2+3}/{r_2-1} \approx 1$ and $|{4 r_1}/{r_2 - 1}| \ll 1$. 
Using Eqs.~\eqref{eq:hierarchy} and \eqref{eq:r_1r_2}, one can further derive approximate expressions for $y_\mu$ and $y_e$:
\be
\ba
\tilde{y}_\mu &\approx \tilde{y}_s + \tilde{y}_b \sin^2 \theta_{23}^q (1 - c_e^2) \sin^2 a_2 \,, \\
\tilde{y}_e &\approx
\left[ \tilde{y}_d + \tilde{y}_s \sin^2 \theta_{12}^q + \tilde{y}_b \sin^2 \theta_{13}^q (1 - c_e^2) \sin^2 a_1' \right] \\
& - \frac{y_b^2}{\tilde{y}_\mu} \sin^2 \theta_{13}^q \sin^2 \theta_{23}^q 
\left[ (\cos a_1' \cos a_2 + c_e^2 \sin a_1' \sin a_2)^2 + c_e^2 \sin^2 (a_1' - a_2) 
\right] \,,
\ea
\ee
where $a_1' = a_1 - \delta^q$. 
However, the relation shown in Eq.~\eqref{eq:r_1r_2} is not always satisfied as a few points with $r_2 \sim 0.13$ are also allowed by $\chi^2 <10$ in our scan and this leads to $({r_2+3})/({r_2-1}) \approx -3.6$. 

The bottom-right panels of \figsref{fig:chisq_10_octant1}{fig:chisq_10_octant2} show the prediction of the lightest neutrino mass, $m_1$, and the effective mass parameter defined as:
\be
m_{\beta\beta}= \Big|\sum_{i=1}^{3} m_i (U_{\rm PMNS})_{e i}^2 \Big|\,,
\ee
which neutrinoless double-$\beta$ decay experiments can determine if neutrinos are Majorana as predicted by our model.
$m_1$ is predicted to be less than $2$~meV and $m_{\beta\beta} \sim (2,20)$~meV. 
The above numerical analysis applies to only the normal ordering of light neutrino masses. We completed a preliminary scan for the inverted ordering and found that very few points had $\chi^2 <100$ and did not proceed with a more dense scan. Although we have not proved that the GUT model disagrees with the inverted mass ordering, our model shows a preference for normal mass ordering given our scan. 

\subsection{The benchmark study}
In this subsection, we study BP1 focusing on the flavour sector of quarks and leptons. Inputs and predictions of fermion Yukawas and mixing parameters are shown in \tabref{tab:benchmark}
\begin{table}[tbp] 
\begin{center}
\begin{tabular}{ l |c c c c c c}
\hline \hline 
Inputs & $a_1$ & $a_2$ & $c_\nu$ & $m_0$ & \hspace{-2mm}$(\eta_u, \eta_c, \eta_t; \eta_d, \eta_s, \eta_b)$\\ 
 & $63.57^\circ$ & $84.17^\circ$ & -1.945 & 82.82\,meV & \hspace{-2mm}$(+,+,-; +,-,+)$ \\\hline
Outputs & $\theta_{13}$ & $\theta_{12}$ & $\theta_{23}$ & $\delta$ & $m_1$ \\
 & $8.53^\circ$ & $32.7^\circ$ & $41.9^\circ$ & $-125^\circ$ & $3.36$ meV \\\cline{2-6}
 ($\chi^2=0.33$)&  $m_{\beta\beta}$ & & $M_{N_1}$ & $M_{N_2}$ & $M_{N_3}$ & \\
& 5.83\,meV & & \!$4.23 \cdot 10^{11}$\,GeV\! & \!$5.32 \cdot 10^{11}$\,GeV\! & \!$1.66 \cdot 10^{13}$\,GeV\! \\\cline{2-6}
\hline \hline
\end{tabular}
\end{center}
\caption{Inputs and predictions of neutrino masses and mixing parameters of BP1 fully satisfy all experimental data. Charged fermion masses and CKM mixing are all fixed at experimental best-fit values. Neutrino masses with normal ordering are predicted. 
\label{tab:benchmark}}
\end{table}
from which we obtain
\begin{alignat*}{2} \label{eq:benchmark_Yukawa}
& h =10^{-2} \left(
\begin{array}{ccc}
 -0.1934 & 0.1343 & -0.0845 \\
 0.1343 & 0.3924 & -0.0995 \\
 -0.0845 & -0.0995 & -33.7016 \\
\end{array}
\right)\,, \nonumber\\
& f = 10^{-2} \cdot \left(
\begin{array}{ccc}
 -0.1934 & 0.1343 & -0.0845 \\
 0.1343 & 0.3924 & -0.0995 \\
 -0.0845 & -0.0995 & -33.7016 \\
\end{array}
\right)\,, \\
& h^\prime = 10^{-2} \cdot \left(
\begin{array}{ccc}
 0. & -0.0693 & 0.0025 \\
 0.0693 & 0. & -1.4430 \\
 -0.0025 & 1.4430 & 0. \\
\end{array}
\right)\,.
\end{alignat*}
From this, the Yukawa and neutrino mass matrices are obtained:
\be
\label{eq:benchmark1}
Y_u = \left(
\begin{array}{ccc}
 2.54\cdot 10^{-6} & 0 & 0 \\
 0 & -0.00137 & 0 \\
 0 & 0 & -0.428 \\
\end{array}
\right)\,,
\ee
\be
Y_d = 10^{-2} \cdot \left(
\begin{array}{ccc}
 0.0056 & -0.0039+0.0014 i & 0.0024\, -0.0 i \\
 -0.0039-0.0014 i & -0.0100 & 0.0029\, +0.0281 i \\
 0.0024\, +0.0 i & 0.0029\, -0.0281 i & 0.5686 \\
\end{array}
\right)\,,
\ee
\be
Y_e = 10^{-2} \cdot \left(
\begin{array}{ccc}
 -0.0018 & 0.0012\, -0.0111 i & -0.0008+0.004 i \\
 0.0012\, +0.0111 i & -0.0003 & -0.0009-0.2304 i \\
 -0.0008-0.0004 i & -0.0009+0.2304 i & 0.9155 \\
\end{array}
\right)
\,,\ee
\be\label{eq:Yukneut}
Y_\nu = 10^{-2} \cdot \left(
\begin{array}{ccc}
 -0.7743 & 0.5374\, +0.1348 i & -0.3379-0.0049 i \\
 0.5374\, -0.1348 i & 1.1586 & -0.3979+2.8068 i \\
 -0.3379+0.0049 i & -0.3979-2.8068 i & -6.4066 \\
\end{array}
\right)\,,\ee
\be
M_\nu = 10^{-2} \cdot \left(
\begin{array}{ccc}
 -0.5269+0.0 i & 0.3628\, +0.0090 i & -0.0434-0.0446 i \\
 0.3628\, +0.0090 i & 0.7407\, -0.0058 i & -0.3755-2.417 i \\
 -0.0434-0.0446 i & -0.3755-2.4168 i & -2.9181+2.0125 i \\
\end{array}
\right)~{\rm eV} \,.
\ee
Diagonalisation of $Y_e$ and $M_\nu$ gives rise to the lepton masses and mixing, and the above benchmark provides 
 $\chi^2 =0.33$. We show predictions of charged lepton Yukawa couplings, mixing angles and the Dirac CP-violating phase, as well as the lightest neutrino mass $m_1$ and $m_{\beta\beta}$ 
in \tabref{tab:benchmark}. 
Applying the type-I seesaw formula, the right-handed neutrino mass matrix is
\begin{eqnarray}
M_{\nu_R} &=& 10^{13} \cdot \left(
\begin{array}{ccc}
 -0.0354 & 0.0246 & -0.0154 \\
 0.0246 & 0.0467 & -0.0182 \\
 -0.0154 & -0.0182 & 1.6650 \\
\end{array}
\right)~{\rm GeV}\,.
\end{eqnarray}
Moreover, the three associated eigenvalues are given in \tabref{tab:benchmark}. Finally, we note that the heaviest right-handed neutrino mass is given by $1.6\times 10^{13}$~GeV. This is lower than the lowest intermediate scale $M_1$ and thus consistent with proton decay measurement.

\section{Leptogenesis}\label{sec:leptogenesis}
As $SO(10)$ GUTs predict very massive right-handed neutrinos, leptogenesis is a natural consequence of such a framework. $SO(10)$ leptogenesis was initially studied in \cite{Branco:2002kt,Nezri:2000pb}. Later, the importance of flavour effects was emphasised \cite{DiBari:2008mp} and in \cite{DiBari:2010ux} the same authors investigated the correlations between the viable $SO(10)$ leptogenesis parameter space and low-scale observables. Leptogenesis within a specific $SO(10)$ model, which fit low energy SM fermionic data \cite{Dueck:2013gca}, was investigated in \cite{Fong:2014gea}. Fitting leptogenesis together with all the fermion mass observables by solving density matrix equations was carried out in \cite{Mummidi:2021anm}. In this section, we calculate the baryon asymmetry produced via thermal leptogenesis for the points of the parameter space scan, which are consistent with the quark and lepton experimental data with $\chi^2<10$ and later emphasise the connection between $SO(10)$ leptogenesis and observables such as proton decay and GWs. 

To calculate the associated baryon-to-photon ratio, we solve the Boltzmann equations
which determine the time evolution of the lepton asymmetry that manifests from the CP-violating and out-of-equilibrium decays of the right-handed neutrinos and associated washout processes. In the simplest formulation, these kinetic equations are in the one-flavoured regime, in which only a single flavour of charged lepton is accounted for. This regime is only realised at sufficiently high temperatures ($T\gg 10^{12}$ GeV) when the rates of processes mediated by the charged lepton Yukawa couplings are out of thermal equilibrium. Therefore, a single charged lepton flavour state is a coherent superposition of the three flavour eigenstates. However, if leptogenesis occurs at lower temperatures ($10^{9}\ll T \ll10^{12}$ GeV), 
scattering induced by the tau Yukawa couplings can cause the single charged lepton flavour to decohere, and the dynamics of leptogenesis must be described in terms of two flavour eigenstates. In such a regime, a density matrix formalism \cite{Barbieri:1999ma, Abada:2006fw,DeSimone:2006nrs,Blanchet:2006ch,Blanchet:2011xq} allows for a more general description than the one-flavoured semi-classical Boltzmann equations. We begin by rotating the Yukawa coupling of the leptonic and Higgs doublet to the right-handed neutrinos mass basis. For example, the benchmark case in \equaref{eq:Yukneut} is rotated to 
 \be
\widetilde{Y}_\nu = 
10^{-2} \cdot \left(
\begin{array}{ccc}
 0.0547\, +0.9061 i & 0.2923\, -0.2626 i & 0.1159\, -0.1146 i \\
 -0.0024+0.04351 i & -1.8277+0.1813 i & -0.4079+1.2977 i \\
 -0.7770-0.2221 i & 0.5467\, +2.3425 i & -6.8722-0.0676 i \\
\end{array}
\right)\,.
\ee
The CP-asymmetry matrix, describing the decay asymmetry generated by $N_{i}$ is denoted by $\epsilon^{(i)}_{\alpha\beta}$, and may be written as \cite{Covi:1996wh}:
\be
\ba
 \varepsilon_{\alpha \beta}^{(i)}&=\frac{3}{32 \pi\left(\tilde{Y}_\nu^{\dagger} \tilde{Y}_\nu\right)_{i i}} \sum_{j \neq i}\left\{i \left[\tilde{{Y}_\nu}_{\alpha i} \tilde{{Y}_\nu}_{\beta j}^{\star}\left(\tilde{Y}_\nu^{\dagger} \tilde{{Y}_\nu}\right)_{j i}-\tilde{{Y}_\nu}_{\beta i}^{\star} \tilde{{Y}_\nu}_{\alpha j}\left(\tilde{{Y}_\nu}^{\dagger} \tilde{{Y}_\nu}\right)_{i j}\right] \frac{\xi\left(x_{j} / x_{i}\right)}{\sqrt{x_{j} / x_{i}}}\right.\\
&\left.+i \frac{2}{3\left(x_{j} / x_{i}-1\right)}\left[\tilde{{Y}_\nu}_{\alpha i} \tilde{{Y}_\nu}_{\beta j}^{\star}\left(\tilde{{Y}_\nu}^{\dagger} \tilde{{Y}_\nu}\right)_{i j}-\tilde{{Y}_\nu}_{\beta i}^{\star} \tilde{{Y}_\nu}_{\alpha j}\left(\tilde{{Y}_\nu}^{\dagger} \tilde{{Y}_\nu}\right)_{j i}\right]\right\},
\ea
\ee
where $x_{i} \equiv M^2_{N_{i}} / M_{N_{1}}^{2}$ and Greek and Roman indices denote charged lepton flavour and right-handed neutrino generation indices, respectively, and 
\begin{equation}
 \xi(x)=\frac{2}{3} x\left[(1+x) \ln \left(\frac{1+x}{x}\right)-\frac{2-x}{1-x}\right]\,.
 \end{equation}
 $N^{B-L}$ is a density matrix parametrising the lepton asymmetries in flavour space, and $\mathcal{P}^{(i)0} $
denotes the projection matrices which describe how a given flavour of lepton is washed out:
\begin{equation}
 N^{B-L} = \left(\begin{array}{ccc}
 N_{\tau\tau} & N_{\tau\mu} & N_{\tau e}\\
 N_{\mu\tau} & N_{\mu\mu} & N_{\mu e}\\
 N_{e\tau} & N_{e \mu} & N_{e e}\\
 \end{array}\right)\,,
\quad
\mathcal{P}^{(i)0} = \frac{1}{\left(\tilde{{Y}_\nu}^{\dagger} \tilde{{Y}_\nu}\right)_{ii}}\left(\begin{array}{ccc}\left|\tilde{{Y}_\nu}_{\tau i}\right|^{2} & \tilde{{Y}_\nu}_{\tau i} \tilde{{Y}_\nu}_{\mu i}^{\star} & \tilde{{Y}_\nu}_{\tau i} \tilde{{Y}_\nu}_{e i}^{\star} \\ \tilde{{Y}_\nu}_{\tau i}^{\star} \tilde{{Y}_\nu}_{\mu i} & \left|\tilde{{Y}_\nu}_{\mu i}\right|^{2} & \tilde{{Y}_\nu}_{\tau i}^{\star} \tilde{{Y}_\nu}_{e i} \\ \tilde{{Y}_\nu}_{e i} \tilde{{Y}_\nu}_{\tau i}^{\star} & \tilde{{Y}_\nu}_{\mu i} \tilde{{Y}_\nu}_{\tau i}^{\star} & \left|\tilde{{Y}_\nu}_{e i}\right|^{2} \end{array}\right)\,.
\end{equation}
 Finally, the density matrix equations which track the time evolution of the lepton asymmetry generated by the decays and washout of the three right-handed neutrinos are given by: 
\be\label{eq:DME}
\ba
\frac{d N_{\alpha \beta}^{B-L}}{d z} = \sum_{i=1}^{3}\varepsilon_{\alpha \beta}^{(i)}& D_{i}\left(N_{N_{i}}-N_{N_{i}}^{\mathrm{eq}}\right)-\frac{1}{2} W_{i}\left\{\mathcal{P}^{(i) 0}, N^{B-L}\right\}_{\alpha \beta} \nonumber\\ 
-& \frac{\operatorname{Im}\left(\Lambda_{\tau}\right)}{H z}\left[\left(\begin{array}{lll}1 & 0 & 0 \\ 0 & 0 & 0 \\ 0 & 0 & 0\end{array}\right),\left[\left(\begin{array}{lll}1 & 0 & 0 \\ 0 & 0 & 0 \\ 0 & 0 & 0\end{array}\right), N^{B-L}\right]\right]_{\alpha \beta} \nonumber \\ 
-&\frac{\operatorname{Im}\left(\Lambda_{\mu}\right)}{Hz}\left[\left(\begin{array}{lll}0 & 0 & 0 \\ 0 & 1 & 0 \\ 0 & 0 & 0\end{array}\right),\left[\left(\begin{array}{lll}0 & 0 & 0 \\ 0 & 1 & 0 \\ 0 & 0 & 0\end{array}\right), N^{B-L}\right]\right]_{\alpha \beta}\,,
\ea
\ee
\begin{figure}[t!]
\centering
\includegraphics[width=\textwidth]{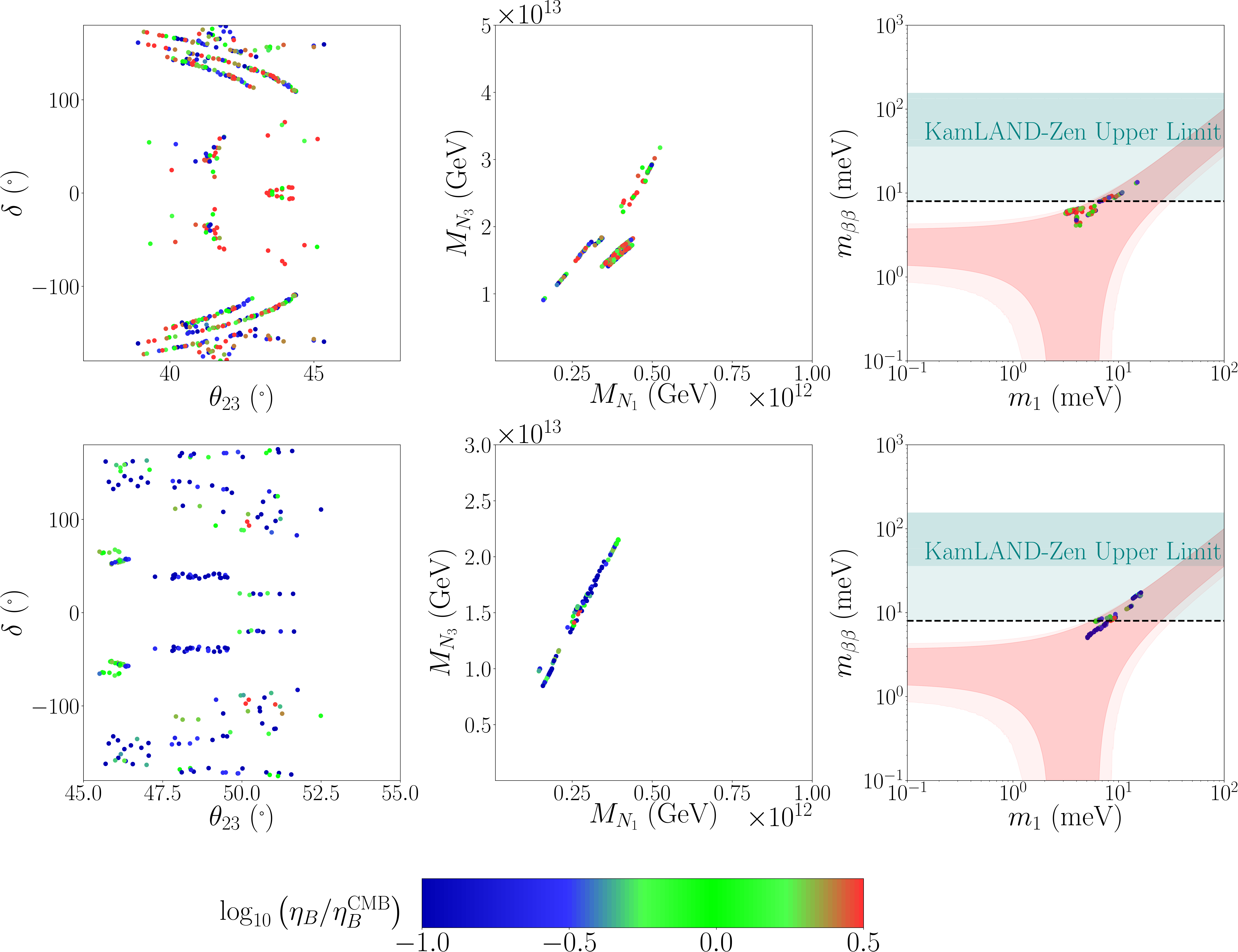}
\caption{The top (bottom) left and centre panels are the two-dimensional correlations between predicted observables for $\chi^2<10$ and $\theta_{23}\leq 45^{\circ}$ ($\theta_{23}\geq 45^{\circ}$). The top (bottom) rightmost panel shows the predictions for the effective neutrino mass for $\theta_{23}\leq 45^{\circ}$ ($\theta_{23}\geq 45^{\circ}$). The colour of the points denotes the ratio of the predicted baryon-to-photon ratio to the experimentally observed best-fit value as measured using CMB data $\eta_B^{\rm CMB} = 6.15 \times 10^{-10}$ \cite{Planck:2015fie}. Consistency with gauge unification is considered. In the leftmost plots, the dashed line labels the sensitivity of the next generation experiments on $0\nu \beta \beta$ decay.}
\label{fig:lepto_octant1}
\end{figure}
where $z=M_{N_{1}}/T$ is the evolution parameter, $H$ is the Hubble expansion rate, and the decay and washout terms are given by 
\be
D_{i}(z)=K_{i} x_{i} z \frac{\mathcal{K}_{1}\left(z_{i}\right)}{\mathcal{K}_{2}\left(z_{i}\right)}\,, \quad W_{i}(z)=\frac{1}{4} K_{i} \sqrt{x_{i}} \mathcal{K}_{1}\left(z_{i}\right) z_{i}^{3}\,,
\ee
where $z_{i} \equiv \sqrt{x_{i}} z$, $\mathcal{K}_{1}$ and $\mathcal{K}_{2}$ are modified Bessel functions of the second kind with the decay asymmetry $K_{i}$ given by
\be
K_{i} \equiv \frac{\tilde{\Gamma}_{i}}{H\left(T=M_{N_{i}}\right)}, \quad \tilde{\Gamma}_{i}=\frac{M_{N_{i}}\left(\widetilde{Y}_\nu^{\dagger} \widetilde{Y}_\nu\right)_{i i}}{8 \pi}\,,
\ee
respectively. The thermal widths of the charged leptons, $\Lambda_{\tau}$, $\Lambda_{\mu}$, are given by the imaginary part of the self-energy correction to the charged lepton propagators in the plasma, and this mediates flavour correlations (for further details of these equations and their application we refer the reader to Ref.~\cite{Moffat:2018wke}).

For each point in our parameter scan, we have solved \equaref{eq:DME} which provides the baryon-to-photon ratio using the publicly available tool {\tt ULYSSES} \cite{Granelli:2020pim} and the associated ``3DME'' code which accounts for the decays and washout of all three right-handed neutrinos. 
We show our results in \figref{fig:lepto_octant1} where all the points have $\chi^2 <10$ and satisfy the proton decay bounds. We observe that for both octants, many low $\chi^2$ points achieve thermal leptogenesis successfully, and their baryon-to-photon ratio is $\eta_B \approx 10^{-10}$. Interestingly, the predicted baryon-to-photon ratio shows little dependence on $\delta$ but has a very
constrained prediction for the effective Majorana mass,  $ 4\lesssim m_{\beta\beta}\,(\text{meV}) \lesssim 10$. This indicates that the predicted Majorana phases are highly constrained within our model.

For all the points with $\chi^2<10$, we found leptogenesis is always in the strong washout regime $K_{1}\gg 1$
 (for the benchmark point $K_1\approx 130$) since the Yukawa couplings (\equaref{eq:Yukneut}) are not very small. As we are in the strong washout regime, neglecting finite density effects is a good approximation \cite{Beneke:2010wd}. 
For the above benchmark point, the lightest two masses of right-handed neutrinos are $M_{N_1} = 4.23 \cdot 10^{11}$~GeV and $M_{N_2} = 5.32 \cdot 10^{11}$~GeV and the baryon-to-photon ratio is $\sim 6.11 \times 10^{-10}$).
Next generation $0\nu \beta \beta $ experiments such as Legend-1000 \cite{https://doi.org/10.48550/arxiv.2107.11462}, nEXO \cite{Adhikari:2021}, NEXT-HD \cite{https://doi.org/10.48550/arxiv.1906.01743}, DARWIN\cite{https://doi.org/10.48550/arxiv.2003.13407}, SNO+II \cite{Andringa_2016} and CUPID-Mo \cite{Armengaud_2020} will allow to test further test the results of our scan. The most sensitive experiments will probe $m_{\beta \beta}$ down to $8-10 \rm{meV}$ which, as we can see from \figref{fig:lepto_octant1}, covers a significant fraction of the points of our scan.

\section{Testability of $SO(10)$ GUT and leptogenesis using gravitational waves}\label{sec:testability}
The creation of topological defects in GUT symmetry breaking is ubiquitous \cite{Jeannerot:2003qv} and in 
our breaking chain produces unwanted defects in the breaking of $SO(10)\to G_2$ (monopoles and domain walls), but in the final breaking, $G_1 \to G_{\rm SM}$, cosmic strings are formed. We assume that inflation occurs after the unwanted defects are formed but before string formation.\footnote{Studies on inflated cosmic strings are discussed in Refs.~\cite{King:2020hyd,Lazarides:2021uxv,Lazarides:2022jgr,Maji:2022jzu}.} Under this assumption, the string network can intersect to form loops which oscillate and emit energy gravitational radiation that can constitute a stochastic gravitational wave background. Importantly, this background can, in principle, be observed by currently running and future GW experiments. 

We assume the Nambu-Goto string approximation, where the string is infinitely thin with
no couplings to particles \cite{Vachaspati:1984gt}, and 
 the amplitude of the relic GW density parameter is:
\be
\Omega_{\rm GW} (f) = \frac{1}{\rho_c} \frac{d\rho_{\rm GW}}{d \log f}\,,
\ee
where $\rho_c$ is the critical energy density of the Universe and $\rho_{\rm GW}$ depends on a single parameter, $G\mu$ where $G= M_{\rm pl}^{-2}$ is Newton's constant and  
$\mu$ is the string tension. For strings generated from the gauge symmetry $G_1 = SU(3)_c \times SU(2)_L \times SU(2)_R \times U(1)_X$. 

$G\mu$ is approximately given by \cite{Vilenkin:1984ib}:
\begin{eqnarray}
G\mu \simeq 
\displaystyle\frac{1}{2(\alpha_{2R}(M_1)+\alpha_{1X}(M_1))} \frac{M_1^2}{M_{\rm pl}^2}\,, 
\end{eqnarray}
and hence we can relate the string tension parameter to the lowest intermediate scale of GUT symmetry breaking.
\begin{figure}[t!]
\centering
\includegraphics[width=.85\textwidth]{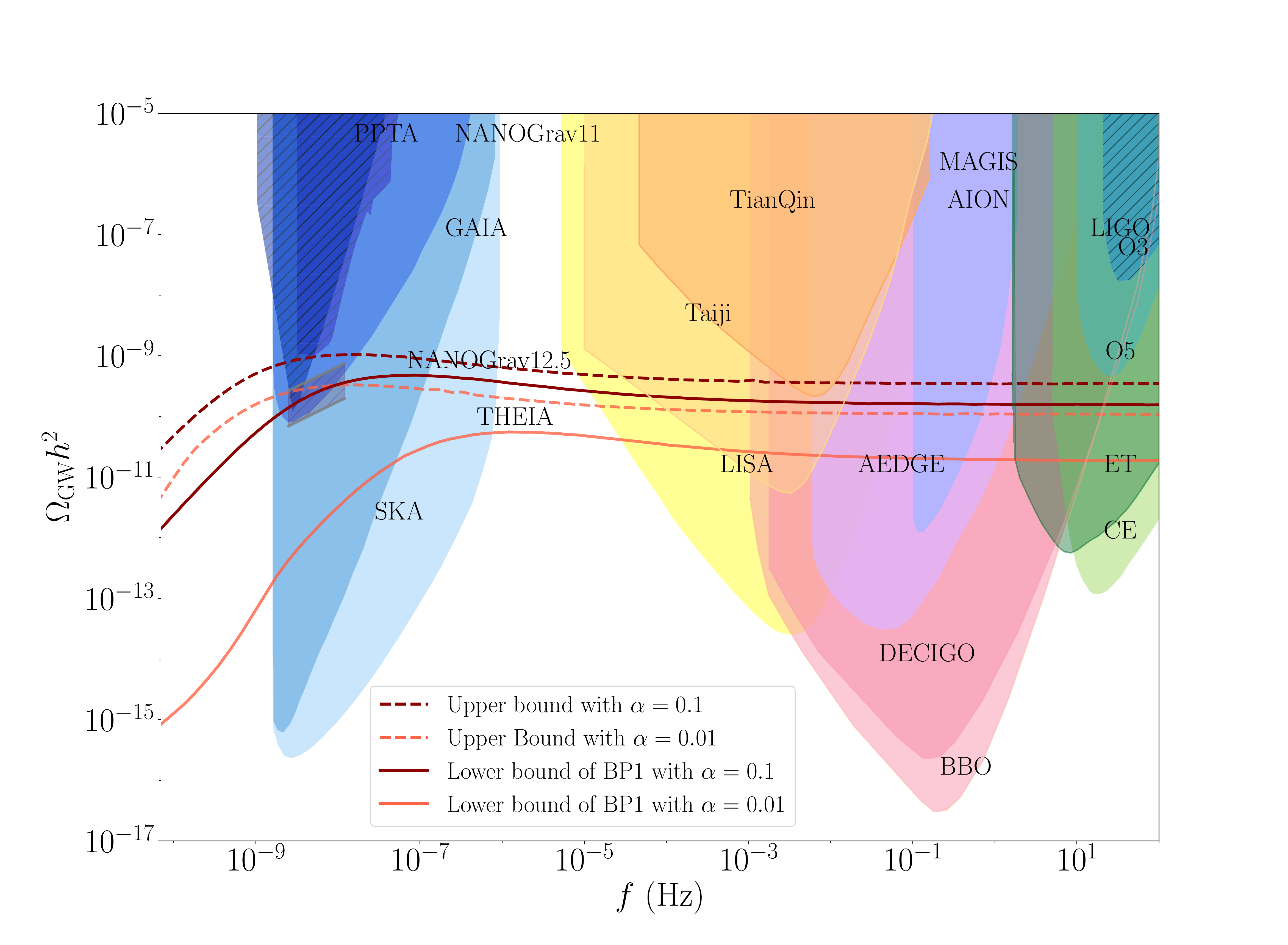}
\caption{Gravitational wave spectrum predicted from the model. Breaking of the intermediate symmetry $G_1 \equiv SU(3)_c\times SU(2)_L \times SU(2)_R \times U(1)_X$ generates cosmic strings with tension $\mu$. The lower bound to the GW spectrum for the benchmark point we considered earlier (red lines) is $G \mu = 2.68 \times 10^{-11}$ corresponding to $M_1 \simeq 2\times 10^{13}$~GeV. The GW spectrums of these two bounds are shown in dashed and solid curves, respectively.}\label{fig:GW}
\end{figure}
The spectrum of $\Omega_{\rm GW} (f) $ is calculated numerically by solving the formulation in \cite{Cui:2018rwi} and \cite{King:2020hyd,King:2021gmj}. A well-known feature is at the high-frequency band, $\Omega_{\rm GW} (f) $ has a $G\mu$-dependent plateau \cite{Blanco-Pillado:2017oxo}. In our calculations, we derive a semi-analytical correlation
\begin{eqnarray}
(\Omega_{\rm GW} h^2)_{\rm plateau} \approx 6.5 \cdot 10^{-6} \times (G\mu)^{1/2} \sim 2.06 \cdot 10^{-5} \times \frac{M_1}{M_{\rm pl}} \,,
\end{eqnarray} 
where numerical values $\alpha_{2R} \sim \alpha_{1X}\sim 1/40$ around the scale $M_1$ have been accounted for. A GW background with $(\Omega_{\rm GW} h^2)_{\rm plateau} \sim 10^{-10}$ is naturally predicted from cosmic strings from the breaking scale $M_1 \sim 10^{13}$~GeV.\footnote{ We note that the non-perturbative production of monopoles can modify the gravitational wave signature from GUT models \cite{Buchmuller:2019gfy}, but this occurs if the monopole and cosmic string generation scales are not well separated. In our GUT model, monopole and string formation are well separated, and we assume a rapid period of inflation between these scales to remove the monopoles.}
There is an important relationship between the string loop length ($l$) at
the time of formation ($t_i$), which is usually taken as a linear relation: $l=\alpha t_i$. 
While $\alpha$ has a distribution of values, it peaks at around 0.1 for both matter and radiation-dominated eras \cite{Blanco-Pillado:2013qja}.

$G\mu$, or equivalently $M_1$, should be restricted to a specific range as its value must be consistent with proton decay measurement and fermion masses and mixing data. For the breaking chain we consider, the upper bound is $M_1^{\rm upper} \simeq 4.4 \times 10^{13}~{\rm GeV}$ from the proton decay. We account for a perturbativity ansatz for Majorana Yukawa coupling, i.e., $M_{N_i} < M_1$ for all right-handed neutrinos.\footnote{The Majorana Yukawa couplings for right-handed neutrinos connect the seesaw scale with the $U(1)_{B-L}$ scale. These couplings, in general, have an upper bound $\lesssim {\cal O}(1)$ to satisfy the perturbativity requirement. To obtain the bound explicitly, one has to consider the RG behaviour of Yukawa couplings and include the influence of Yukawa couplings on gauge unification. This procedure is complicated, and we relegate this for future work. In this work as an alternative, we consider a simplified perturbative ansatz by requiring the heaviest right-handed neutrino mass lighter than the $U(1)_{B-L}$ scale, i.e., $M_{N_3} < M_1$.}
In this ansatz, the lower bound of $M_1$ is obtained as $M_1 > M_1^{\rm lower} \equiv M_{N_3}$.
 This value varies with different scatter points shown in Figs.~\ref{fig:chisq_10_octant1} and \ref{fig:chisq_10_octant2}. For the benchmark point BP1, we take $M_1^{\rm lower} ({\rm BP1})\simeq 2 \times 10^{13}$~GeV and obtain $G\mu \simeq 2.68 \times 10^{-11}$. The associated $\Omega_{\rm GW} (f) h^2$ is shown in \figref{fig:GW} with dashed and solid curves, where the coloured (hatched) regions indicate future (current) experimental sensitivities. 

The prediction of $\Omega_{\rm GW}h^2$ depends on the loop size parameter $\alpha$. While simulations \cite{Blanco-Pillado:2013qja, Blanco-Pillado:2017oxo} show this parameter peaks around $\alpha \approx 1$. Deviation could be considered as there could be some uncertainties in the peak value and distribution around it. A smaller value of $\alpha$ reduces the GW intensity as $\Omega_{\rm GW} \approx \sqrt{\alpha}$. In the plot, we consider the case with the value $\alpha = 0.01$ (light red) as a comparison to the main result derived at $\alpha= 0.1$ (red).

A series of GW observatories, including space-based laser interferometers (LISA \cite{Audley:2017drz}, Taiji \cite{Guo:2018npi}, TianQin \cite{Luo:2015ght}, BBO \cite{Corbin:2005ny}, DECIGO \cite{Seto:2001qf}), atomic interferometers (MAGIS \cite{Graham:2017pmn}, AEDGE \cite{Bertoldi:2019tck}, AION \cite{Badurina:2019hst}), and ground-based interferometers (Einstein Telescope \cite{Sathyaprakash:2012jk} (ET), Cosmic Explorer \cite{Evans:2016mbw} (CE)) will cover GW frequency band from mHz to kHz and probe $G \mu$ values in a wide range $\sim 10^{-19} - 10^{-11}$, as seen in Fig.~\ref{fig:GW}. A combined analysis of these experiments can potentially exclude cosmic strings-originated GW signals to high sensitivity. 

Data from pulsar timing arrays (PTA) such as EPTA \cite{Lentati:2015qwp} and NANOGrav (11-year data set) \cite{Arzoumanian:2018saf} have already probed the nHz regime and provided upper limits through the non-observation of GWs. Future PTA such as SKA \cite{Janssen:2014dka} will cover even more parameter space, and large-scale surveys of stars such as Gaia \cite{Brown:2018dum} and the proposed upgrade, THEIA \cite{Boehm:2017wie}, can be powerful and complementary probes of gravitational waves in the same frequency regime.

\begin{figure}[t!]
\centering
\includegraphics[width=.85\textwidth]{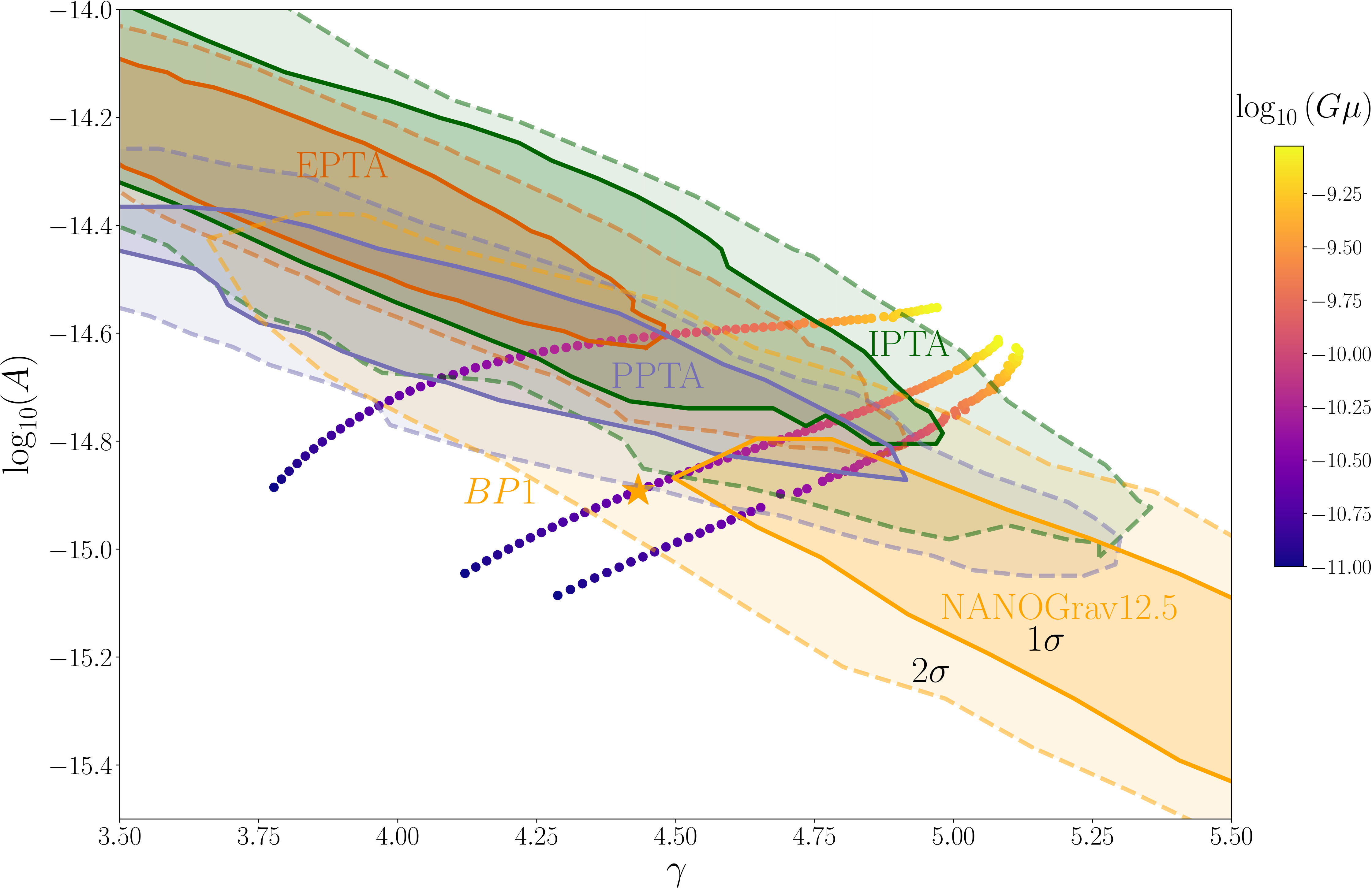}
\caption{Comparison between SGWB signals produced by cosmic strings with $G\mu$ from $10^{-11}$ to $10^{-9}$ the possible $1\sigma$ and $2\sigma$ regions hinted by EPTA, PPTA, IPTA and NANOGrav. The SGWB signal has been fitted for three frequencies: 2.4 nHz, 5.4 nHz and 12 nHz. The simulation shows that the signal is compatible with all experiments at $2\sigma$. The orange star indicates the prediction of $(\gamma, A)$ in BP1.}
\label{fig:curves}
\end{figure}

\begin{figure}[h!]
\centering
\includegraphics[width=.99\textwidth]{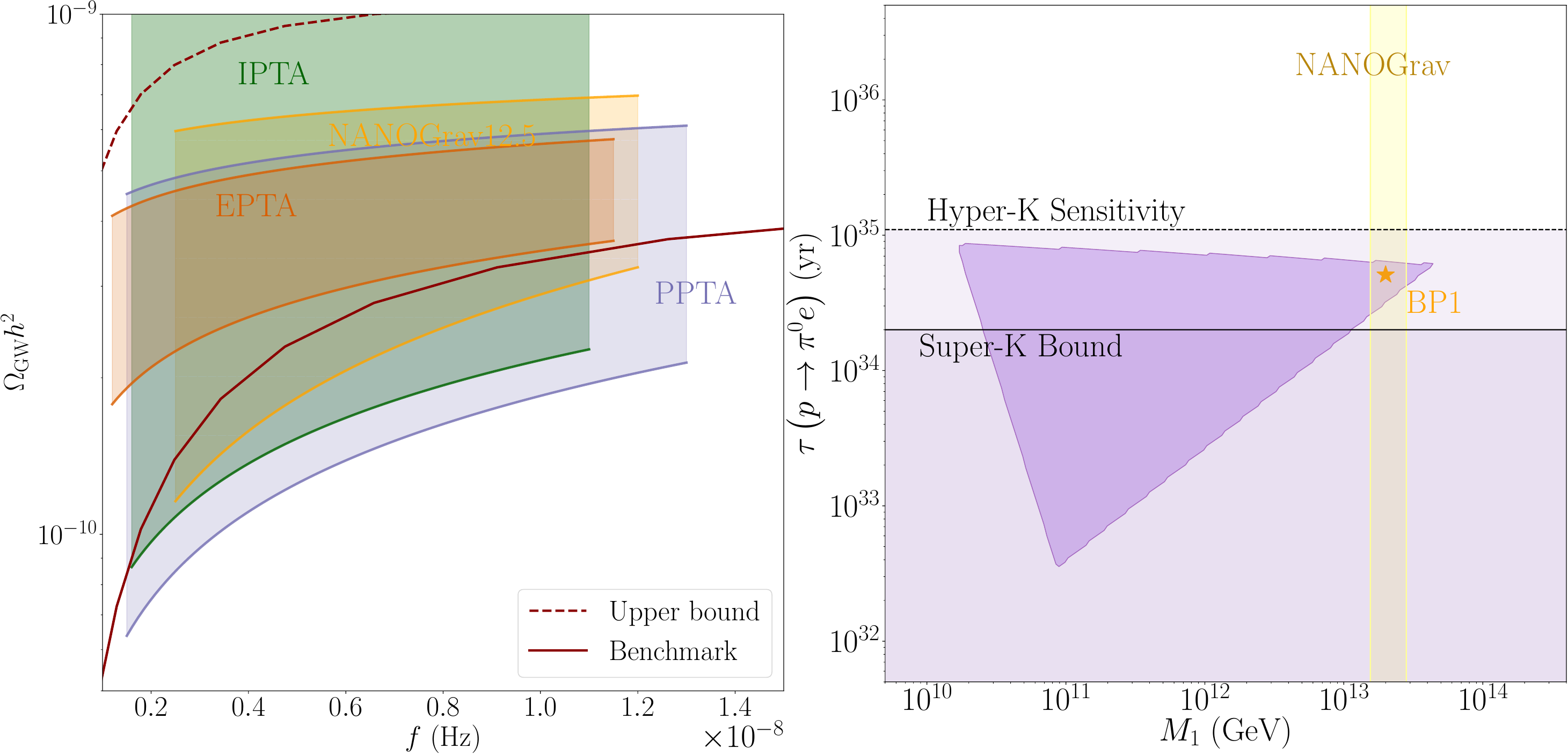}
\caption{Left: Blown-up image of the nHz region of the gravitational waves spectrum of the benchmark point and the upper bound. This is compared with the region of EPTA, IPTA, PPTA and NANOGrav consistent with the observation of an SGWB. Right: Proton decay lifetime compared with the region of M1 consistent with NANOGrav12.5, we can see that there is a region of the parameter space which can be tested by Hyper-K, which is consistent with NANOGrav. The orange star indicates BP1.}
\label{fig:blowup_protondecay}
\end{figure}
The PTA experiment NANOGrav released its 12.5-year data set \cite{Arzoumanian:2020vkk} which announced the detection of a common-spectrum process that has a characteristic strain of the form
\begin{eqnarray}
h_c(f) = A \left( \frac{f}{f_{\rm yr}} \right)^\alpha \,,
\end{eqnarray}
where $f_{\rm yr} = 1 /{\rm year}$, and $A$ is an amplitude parameter referring to the correlation between pulsars. 
If this signal is confirmed as a cosmic GW background, the characteristic strain can be transformed into the GW energy density:
\begin{eqnarray} \label{eq:Omegahsq}
\left(\Omega_{\rm GW}(f) h^2 \right)_{\rm PTA} \approx 2.02\cdot 10^{-10} \left(\frac{A}{10^{-15}}\right)^2 \times \left( \frac{f}{f_{\rm yr}} \right)^{5-\gamma}\,,
\end{eqnarray} 
where $\gamma = 3-2\alpha$. 
Motivated by searches for supermassive black hole mergers, the cross-power spectral density can be fixed at $\gamma = 13/3$ and $A$ as measured by NANOGrav is
\begin{eqnarray}
A = 1.92_{-0.55}^{+0.75} \times 10^{-15}, \quad \gamma = \frac{13}{3} ~\text{fixed\quad (NANOGrav)} 
\end{eqnarray} at 95\% CL. 
This range of $A$ has recently been confirmed by other pulsar-timing arrays including PPTA \cite{Goncharov:2021oub,Chen:2022azo}, EPTA \cite{Chen:2021rqp} and IPTA \cite{Antoniadis:2022pcn} without fixing the value of $\alpha$ or $\gamma$. At 95\% CL, $A$ and $\gamma$ in these experiments are fitted to be
\begin{eqnarray}
&&A = 2.82_{-1.16}^{+0.73} \times 10^{-15} \,, \quad \gamma = 4.11^{-0.41}_{+0.52}\quad ({\rm PPTA}) \,, \nonumber\\
&&A = 5.13_{-2.73}^{+4.20} \times 10^{-15} \,, \quad \gamma = 3.78^{-0.59}_{+0.69}\quad ({\rm EPTA}) \,, \nonumber\\
&&A = 3.8_{-2.5}^{+6.3} \times 10^{-15} \,, \quad \quad \gamma = 4.0^{-0.9}_{+0.9}\qquad ({\rm IPTA}) \,, 
\end{eqnarray}
respectively.
\begin{figure}[t!]
\centering
\includegraphics[width=.99\textwidth]{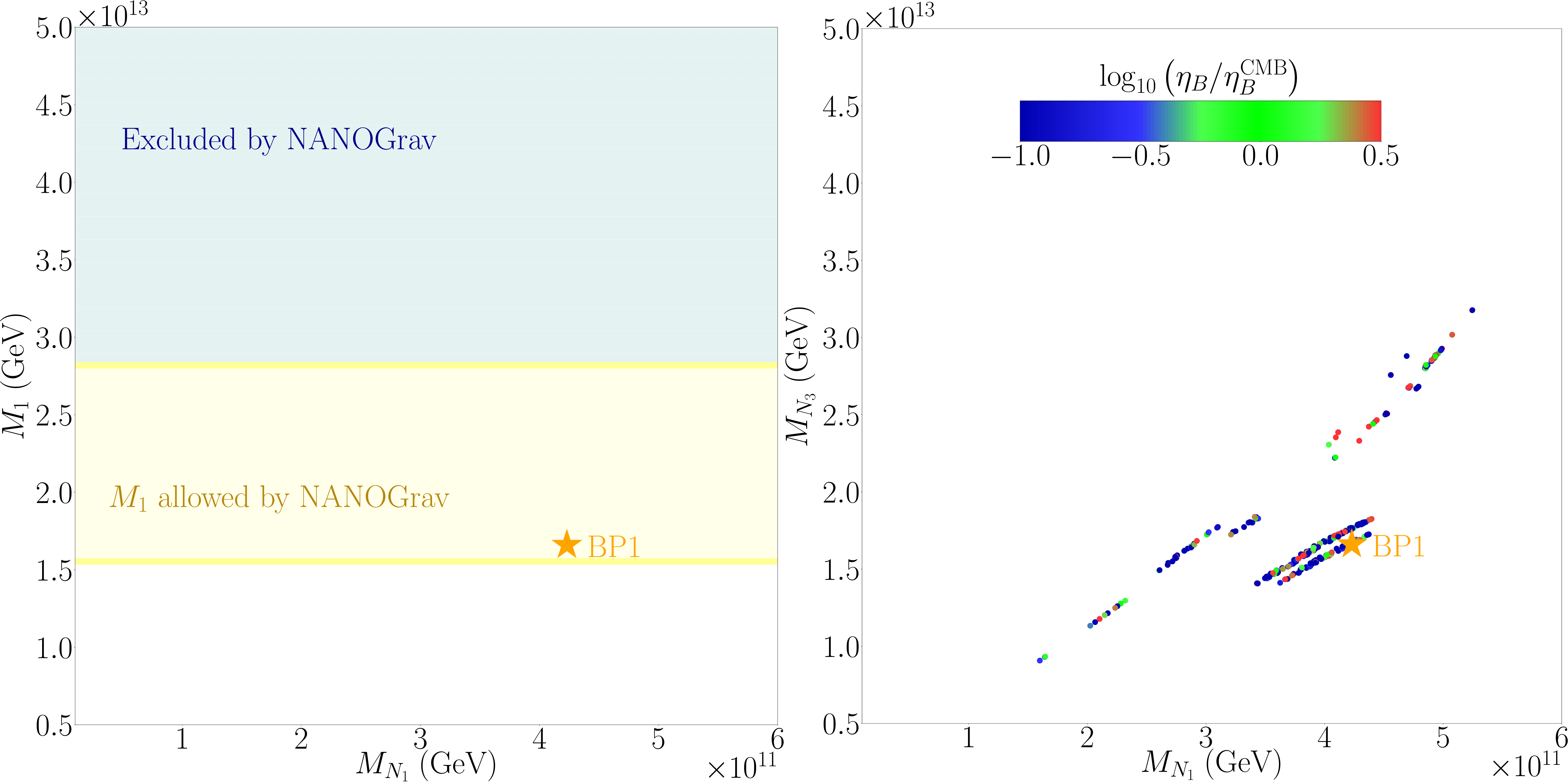}
\caption{Comparison between the results of our scan and NANOGrav12.5 bound. The yellow shaded region includes all the values of $M_1$ consistent with NANOGrav. Applying the perturbativity ansatz, i.e., $M_1>M_{N_3}$, we can see how most of the points in our scan are consistent with NANOGrav. The first octant of lepton mixing angle $\theta_{23}$ is considered.}
\label{fig:RHN_mass_scale}
\end{figure}

These data hint at a GW energy density $\Omega_{\rm GW} h^2 \sim 10^{-10}$ in the nHz band. 
The hinted ranges of $A$ and $\gamma$ are highly correlated in all these experiments. We include $1\sigma$ and $2\sigma$ regions of $(\gamma, A)$ restricted by these measurements in Fig.~\ref{fig:curves}. 
In the same figure, we also present three dotted curves which show how the $G \mu$, associated with cosmic strings, overlays on the $A$-$\gamma$ regions of various experiments. Following the treatment used in \cite{Ellis:2020ena} the curves are obtained as follows: for a given value of $G\mu$, the spectrum of $\Omega_{\rm GW}(f)h^2$ is obtained from our numerical calculation, $A$ and $\gamma$ are reversely calculated via \equaref{eq:Omegahsq} by fixing the frequency $f = f_*$. In particular, $\gamma$ is solved via $\gamma = 5 - \frac{\partial \log\Omega_{\rm GW}(f)}{ \partial \log f}\Big|_{f_*}$. We set $f_* = 2.4, ~5.6$ and $12$~nHz, respectively, vary $G\mu$ from $10^{-11}$ to $10^{-9}$, and obtain three curves respectively, shown in \figref{fig:curves} where the second frequency is the one considered in \cite{Ellis:2020ena} and the first and third refer to the first and the fifth lowest frequencies measured in NANOGrav12.5. The overlap between these curves and the PTA hinted regions indicate
the consistency between cosmic-string-sourced SGWB signal and the experimental searches.
For $f_*=5.6$~nHz, we found that the signal is compatible with NANOGrav12.5, PPTA and IPTA in $1\sigma$ and EPTA
in $2\sigma$ ranges. $A$ and $\gamma$ in $2\sigma$ ranges are restricted in
\begin{eqnarray}
A &=& (1.75,1.95)\times 10^{-15}\,,\quad \gamma = (4.34,4.90) \quad \text{(NANOGrav)}\,, \nonumber\\
A &=& (1.32,1.85)\times 10^{-15}\,,\quad \gamma = (4.44,4.86) \quad \text{(PPTA)}\,, \nonumber\\
A &=& (1.56,1.82)\times 10^{-15}\,,\quad \gamma = (4.68,4.85) \quad \text{(EPTA)}\,, \nonumber\\
A &=& (1.38,2.24)\times 10^{-15}\,,\quad \gamma = (4.50,5.03) \quad \; \text{(IPTA)}\,,
\end{eqnarray}
respectively. Here, upper bounds of $A$ and $\gamma$ in NANOGrav and IPTA are not considered as they predict $\Omega_{\rm GW} h^2 > 10^{-9}$ which are too large for our interests.
We apply these ranges to \equaref{eq:Omegahsq}, and in the left panel of \figref{fig:blowup_protondecay} we 
show a zoom-in of the upper bound and BP1 GW signal intersecting the various PTA experimental sensitivities. We observe that our benchmark point
can explain the possible signal observed. In the right panel of \figref{fig:blowup_protondecay} we show our GUT model's predicted proton lifetime as a
function of $M_{1}$. 

In the left panel of \figref{fig:RHN_mass_scale}, we show the range of $M_1$ values predicted by our model, where the yellow shows the 
region consistent with the NANOGrav 12.5-year data set. As there is a degree of freedom which relates $M_1$ and $M_{N_{1}}$ that we cannot constrain, there is a range of lightest right-handed neutrino masses that can accommodate this GW signal. In the right panel, we show how our leptogenesis prediction 
lies in the $M_{N_{1}}-M_{N_{3}}$ plane.

In \figref{fig:RHN_mass_scale}, we present the plot of leptogenesis vs RHN neutrino masses $M_{N_1}$ and $M_{N_3}$ and include the $M_1$ upper bound required by NANOGrav for the first (left panel) and second (right panel) octant. The heaviest RHN neutrino mass $M_{N_3}$ should not be larger than $M_1$ and thus not larger than the upper bound of $M_1$. 
Importantly, the RHN masses which give viable leptogenesis are in the region of the PS testable by NANOGrav12.5.

Finally, we briefly discuss the GW signal predicted by the benchmark point. Given the lowest and the second lowest intermediate scales $M_1 = 2\times 10^{13}$~GeV and $M_2 = 5 \times 10^{13}$~GeV and the consistency with gauge unification, $G\mu$ is predicted to be $1.1\times 10^{-10}$ which corresponds to $\Omega_{\rm GW}(f) h^2 \sim 4.2 \times 10^{-10}$ at the peak around $6.3 \times 10^{-7}$~Hz. This is a critical value to be tested in the PTA experiments. In the high frequency band, $\Omega_{\rm GW}(f) h^2 \sim 1.6 \times 10^{-10}$ is predicted. The future space-based, atomic and underground-based GW observatories will give an excellent test of this value.

\section{Discussion and conclusion}\label{sec:conclusion}
Grand Unified Theories offer an attractive ultraviolet completion of the Standard Model and can explain the masses and mixings of the known particles. In our former works, we proposed gravitational wave and proton decay measurements as complementary windows to identify possible breaking chains of GUTs \cite{King:2020hyd} and provided a systematic analysis for all $SO(10)$ chains via the Pati-Salam-type breaking. In addition, we focused on the connection between the proton lifetime and stochastic GW background energy density \cite{King:2021gmj}.

In this paper, we perform a detailed analysis of one such SO(10) GUT breaking chain involving a realistic and minimal $SO(10)$ GUT model which breaks to the Standard Model gauge group via three intermediate gauge symmetries. Four Higgs multiplets are required to induce the desired pattern of breaking. An additional two Higgs multiplets are introduced for our model to predict the SM fermion masses and mixing to a high statistical significance. From the constraint of gauge unification and proton decay, and including the relevant particles in our two-loop renormalisation group equations, we determine the scale of spontaneous symmetry breaking from our GUT model to the Standard Model. The final intermediate symmetry breaking
induces a network of cosmic string, which results in the emission of GWs, assuming that
inflation ends earlier than string formation. Our numerical procedure uses the quark masses and mixing as constraints on our GUT model parameter space (four-dimensional). We performed a grid scan of this space and found that neutrino masses and mixing (five observables) can be accommodated to a high statistical significance. Once the viable regions of the model parameter space were found and the intermediate scales determined, we have a fixed prediction for leptogenesis, gravitational waves and proton decay.

Our scan result shows that the right-handed neutrino spectrum can be strongly restricted as the model has to fit all flavour data, including fermion masses, CKM mixing and PMNS mixing. The heaviest right-handed neutrino mass is predicted to be in a narrow region above $10^{13}$~GeV. Given the chosen breaking chain, this region is still consistent with the upper bound of the lowest intermediated scale restricted by proton decay and GW measurements. It is worth emphasising that recently released data from a series of Pulsar Timing Arrays measurements hints at this region. However, more data is required to confirm if these signals are from a stochastic GW background. In the future, both proton decay measurement experiment Hyper-K and GW observatories can exclude or confirm this region. Furthermore, the latter has the potential to provide powerful constraints to many GUT models.

Throughout, we present a benchmark point, BP1, which satisfies all experimental constraints, including proton decay, quark masses and mixing, lepton masses and mixing and can address the observed matter-antimatter asymmetry in the Universe. By fixing intermediate scales at $2\times 10^{13}$ and $5\times 10^{13}$~GeV, respectively, the GUT breaking scale is solved to be around $5.68 \times 10^{15}$~GeV, and the proton lifetime is predicted to be around $5.1\times 10^{34}$ years, which survives the upper bound restricted by Super-K and will be tested in Hyper-K in the future. This benchmark predicts three right-handed neutrino masses $(M_{N_1}, M_{N_2}, M_{N_3}) = (4.23 \cdot 10^{11}, 5.32 \cdot 10^{11}, 1.66 \cdot 10^{13})$~GeV. These heavy neutrinos, together with their CP-violating Yukawa couplings with active neutrinos, can explain the baryon-antibaryon asymmetry of the Universe quantitatively via leptogenesis. The mass of the heaviest right-handed neutrino is lower than the lowest intermediate scale $M_1$, thus consistent with the Super-K constraint on proton decay.

In summary, we are entering an exciting era where the culmination of data from proton decay, gravitational wave and neutrinoless double beta decay
experiments will allow us to test Grand Unification and its connection with the matter-antimatter asymmetry at unprecedented levels. To illustrate this, we have analysed in full detail a specific example of a realistic and minimal SO(10) model broken to the SM via three intermediate scales, and shown that it will be tested very soon by such experiments. Given the upcoming synergy in experimental data it is of interest to extend our  methodology to further GUT models.

\section*{Acknowledgement}
This work was partially supported by the European Union's Horizon 2020 Research and
Innovation Programme under Marie Sklodowska-Curie grant agreement HIDDeN European
ITN project (H2020-MSCA-ITN-2019//860881-HIDDeN), the European Research Council
under ERC Grant NuMass (FP7-IDEAS-ERC ERC-CG 617143). S. F. K. acknowledges
the STFC Consolidated Grant ST/T000775/1. Y. L. Z. was supported by the National Natural Science Foundation of China under grants No. 12205064. B. F. acknowledges the Chinese Scholarship Council (CSC) Grant No.\ 201809210011 under agreements [2018]3101 and [2019]536. This work used the DiRAC@Durham facility managed by the Institute for Computational Cosmology on behalf of the STFC DiRAC HPC Facility (\href{www.dirac.ac.uk}{www.dirac.ac.uk}). The equipment was funded by BEIS capital funding via STFC capital grants ST/P002293/1, ST/R002371/1 and ST/S002502/1, Durham University and STFC operations grant ST/R000832/1. DiRAC is part of the National e-Infrastructure. 
Y. L. Z. would like to thank J. Liu for useful discussion. 
\bibliographystyle{JHEP}
\bibliography{SO10lep}

\end{document}